\documentclass[conference]{IEEEtran}
\IEEEoverridecommandlockouts
\usepackage{cite}
\usepackage{amsmath,amssymb,amsfonts}
\usepackage{algorithmic}
\usepackage{graphicx}
\usepackage{textcomp}
\usepackage{xcolor}
\newcommand{\MyArch}{QCATS\xspace}
\newcommand{\lyy}[1]{{\color{black}#1}}
\usepackage{tabularx}
\usepackage{xcolor} 
\usepackage{multirow}
\usepackage{enumitem}
\usepackage{subcaption}
\usepackage{url}
\usepackage{algorithm}
\usepackage{algorithmic}
\usepackage{booktabs}
\usepackage{graphicx}
\usepackage{xspace}
\usepackage{changes}
\usepackage{makecell} 
\usepackage{amsmath}
\usepackage[utf8]{inputenc}
\usepackage{cite} 
\usepackage{bibunits}
\defaultbibliographystyle{IEEEtran}
\newcommand{\zl}[1]{{\color{black}#1}}
\newcommand{\yy}[1]{{\color{black}#1}}
\def\BibTeX{{\rm B\kern-.05em{\sc i\kern-.025em b}\kern-.08em
    T\kern-.1667em\lower.7ex\hbox{E}\kern-.125emX}}
\begin{document}
\begin{bibunit}
\title{\MyArch: Query Context-Aware Transformer Slicing for Efficient Predictive Query Processing}

\author{
\IEEEauthorblockN{
Yueying Li\textsuperscript{1},
Zhongle Xie\textsuperscript{1,2, *}\thanks{* Corresponding author.},
Ke Chen\textsuperscript{1,2},
Lidan Shou\textsuperscript{1,2}
}
\IEEEauthorblockA{
\textsuperscript{1}The State Key Laboratory of Blockchain and Data Security, Zhejiang University;\\
\textsuperscript{2}Hangzhou High-Tech Zone (Binjiang) Institute of Blockchain and Data Security\\
\{zjuyueyingli, xiezl, chenk, should\}@zju.edu.cn
}
}

\maketitle

\begin{abstract}
\yy{In-database predictive query processing increasingly applies Transformer-based models within relational pipelines.
However, existing in-database inference typically exposes only tuple-level model inputs to the inference runtime, leaving relational predicates and metadata statistics invisible to neural execution planning.}
In this paper, we propose \textbf{\MyArch}, a query context-aware transformer slicing framework that enables efficient sparse inference inside database systems. 
\MyArch executes at query granularity: instead of routing individual tokens or tuples during inference, it uses query predicates and metadata statistics to pre-select context-aligned FFN slices before model execution.
The framework comprises offline expert construction and lightweight query-level routing that dynamically selects experts during execution.
\MyArch further introduces system optimizations, including asynchronous CPU–GPU pipelines and routing-aware batching.
Experiments on four predictive-query workloads with BERT-base and Qwen-0.6B show that \MyArch achieves up to 4.42$\times$ latency reduction while preserving prediction accuracy comparable to dense baselines.
\end{abstract}

\section{Introduction}

\yy{
Modern analytical databases integrate Transformer-based predictive operators into SQL pipelines to process unstructured text alongside relational data. 
Known as in-database predictive querying~\cite{lin2022comparative, peng2025towards}, this paradigm supports semantic analytics through cloud inference APIs or UDFs~\cite{databricks2024serving,google2024inference,alibaba2023polardb}.
These queries typically produce short outputs, such as labels or scores, for many SQL-selected tuples, making accumulated tuple-level inference, rather than autoregressive decoding, a dominant cost of query execution.
}

\begin{figure}[t]
    \centering
    \includegraphics[width=\linewidth]{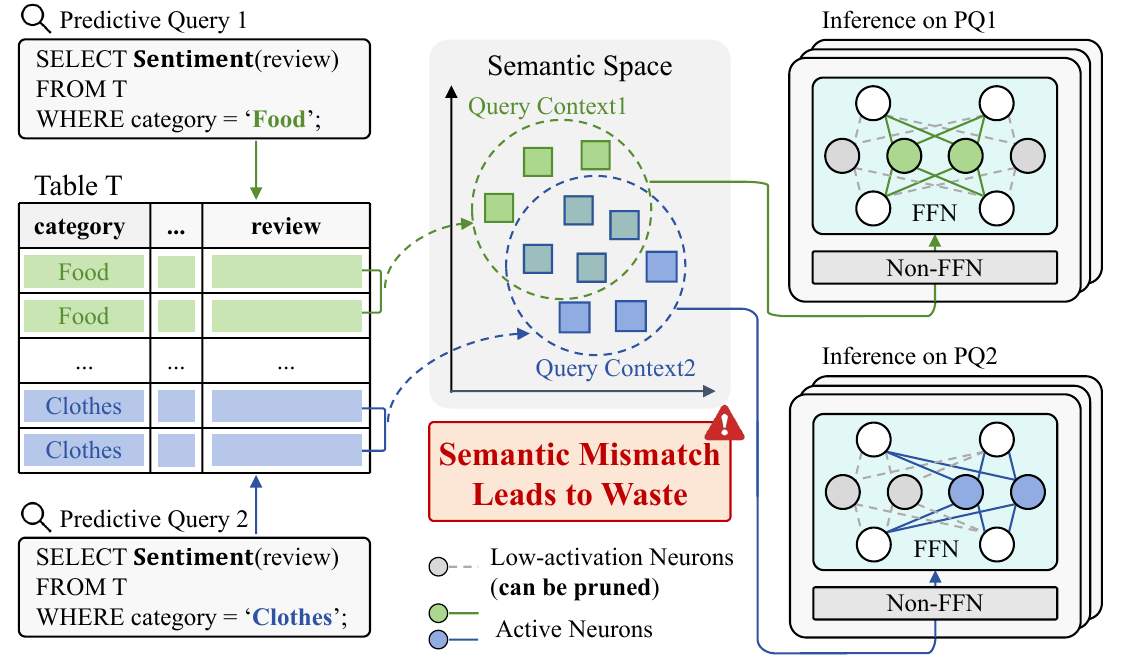}
    \caption{FFN Activation Varies Across Query Contexts.}
    \label{fig:intro}
\end{figure}

\yy{
Opportunities to reduce this cost arise from the structure of predictive queries: inference operates on subsets selected by SQL predicates.
These predicates, together with subset statistics, define a semantic space that often differs from the global data distribution, which we refer to as query context.
}
For Transformer models, different query contexts induce distinct activation patterns in feed-forward networks (FFNs)~\cite{cai2024textit}.
As illustrated in Fig.~\ref{fig:intro}, sentiment analysis over food reviews activates a different subset of neurons from the same task on clothing or electronics reviews, suggesting that only a fraction of model capacity is required for many query workloads. 
\yy{
We further empirically verify this phenomenon in
Fig.~\ref{fig:activation_bert_intro} using eight queries, details are provided in the
appendix.
For each query, we feed its selected text tuples into the fine-tuned
BERT model and compute the average activation of each FFN neuron,
producing one row of the heatmap. Each column represents an FFN neuron,
and darker colors indicate higher average activation.
A small subset of neurons remains highly activated across the queries
and is therefore placed on the left. Beyond this shared subset, highly
activated regions appear at different neuron positions for different
query contexts, indicating that distinct queries activate different
subsets of neurons. This observation motivates selecting
context-aligned FFN slices according to the query context.
}

\begin{figure}[t]
    \centering
    \includegraphics[width=0.5\linewidth]{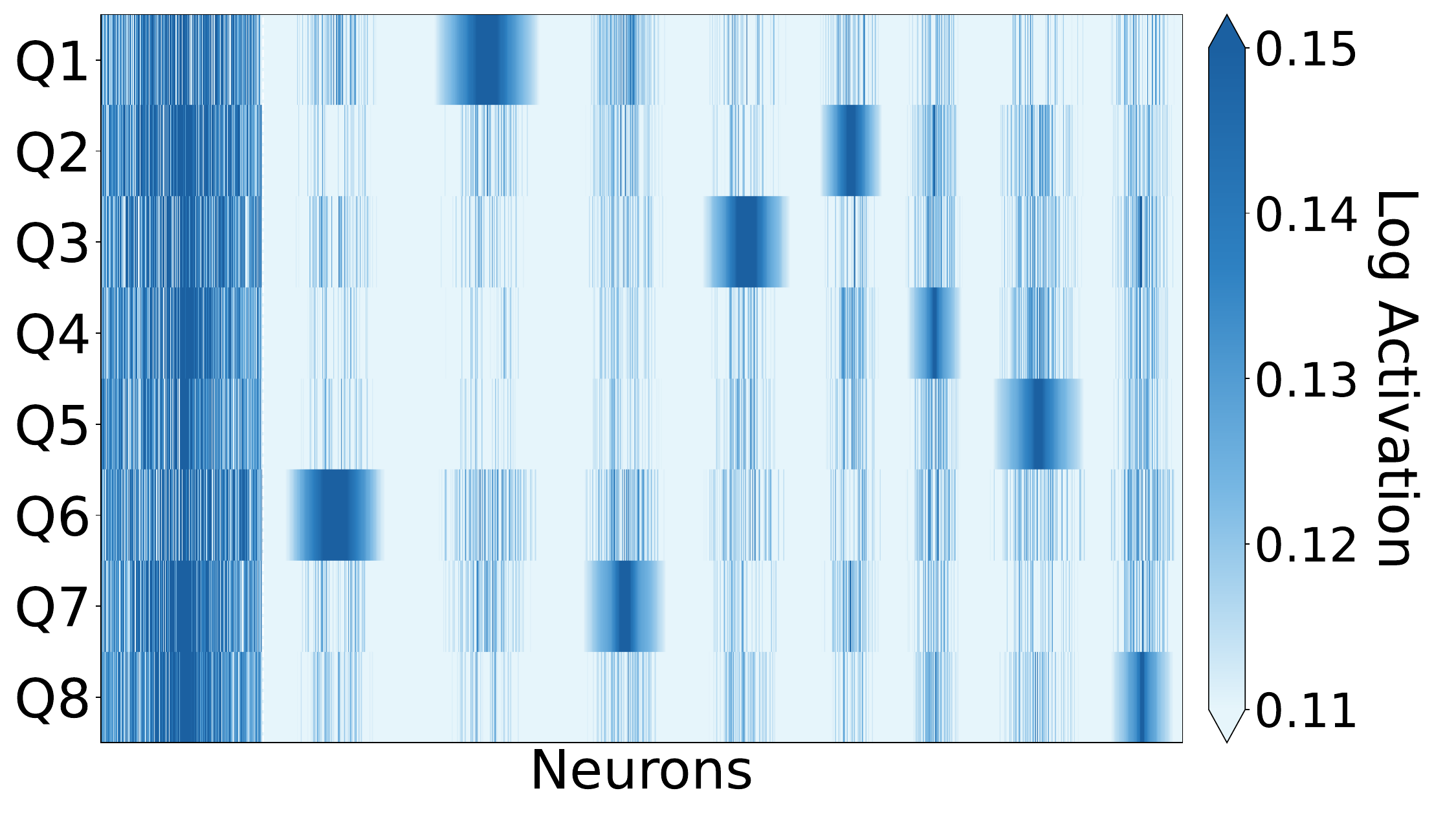}
    \caption{Average FFN neuron activations for text tuples selected by eight queries ($Q1$–$Q8$). }
    \label{fig:activation_bert_intro}
\end{figure}

However, existing database systems typically execute inference through black-box user-defined functions (UDFs)~\cite{databricks2024serving,google2024inference,alibaba2023polardb}, transmitting only raw text data to the model \zl{while} discarding relational constraints and metadata statistics.
Even when the inference engine applies generic batching or kernel-level acceleration, these optimizations are unaware of the relational context that determines which tuples are semantically similar.
Consequently, models operate in a context-agnostic manner and execute dense inference for every input tuple. 
This mismatch leads to substantial waste in computation and memory resources in predictive query processing. 
That is, the computation repeatedly activates the full model despite the query-level locality of predictive workloads.

From a systems perspective, the inefficiency motivates conditional execution, where only computation relevant to the current query context is activated during inference.
Analogous to predicate pushdown~\cite{park2022end,yan2023predicate,guo2025database}  and column pruning~\cite{abadi2008column, jose2023multilingual}, which avoid irrelevant data movement and attribute access, query-aware neural execution should avoid irrelevant parameter access and activation computation.
Unfortunately, existing neural inference pipelines lack mechanisms to translate query semantics into deterministic and database-friendly conditional model execution.

Sparse activation techniques like Mixture-of-Experts (MoE)~\cite{lepikhin2020gshard,fedus2022switch} partially address this issue by enabling conditional computation within neural networks. 
\zl{
However, they rely on token-level, layer-wise routing during inference, which introduces control-flow divergence and dynamic data redistribution overheads, degrading GPU utilization~\cite{he2021fastmoe,gale2023megablocks}.
}
In database workloads, this overhead is amplified because tuples selected by the same query are naturally processed in large batches; token-level routing breaks this homogeneity and prevents the executor from constructing a single static inference path for the batch.
Furthermore, \zl{
it \lyy{introduces additional buffering overhead}
for fragmented tokens across experts, causing inflated activation memory and unpredictable resource demands.
}

Despite extensive research on sparse inference and expert routing, existing approaches~\cite{hellerstein2012madlib,boehm2020systemds,jasny2020db4ml} are primarily designed for standalone model serving and do not exploit the semantic structure exposed by database queries. 
To our knowledge, no prior system leverages query-level context to enable deterministic, database-integrated sparse execution of \zl{Transformer} models.
This gap prevents current solutions from achieving both high inference efficiency and tight integration with relational execution pipelines.

Beyond model-level limitations, predictive query processing also introduces system-level challenges.
CPU-side preprocessing, including tokenization and data preparation, can become a bottleneck when GPU inference is accelerated on-the-fly. 
Meanwhile, long-tailed text length distributions~\cite{zheng2025bucketserve} cause excessive padding overhead under conventional batching, while synchronous execution pipelines provide insufficient CPU--GPU overlap, leading to pipeline stalls and underutilized hardware resources.

In this work, we propose \uline{Q}uery \uline{C}ontext-\uline{A}ware \uline{T}ransformer \uline{S}licing (\MyArch), a framework that rethinks sparse inference for \yy{Transformer-based} models from a database-first perspective.
Instead of performing routing inside the model, \MyArch shifts routing decisions to the query level.
By leveraging the query context, \MyArch pre-determines the subset of model parameters that will be activated before inference begins.
This design enables static, deterministic routing, avoids runtime control-flow divergence, and aligns sparse execution with the granularity of database workloads.
In this sense, QCATS treats model inference as a query physical operator whose execution path can be specialized using optimizer-visible relational context before tuple-level inference begins.

\MyArch consists of two tightly integrated components. 
First, \zl{it transforms} a fine-tuned dense model into a library of context-aligned expert slices by exploiting FFN activation sparsity, enabling sparse inference without retraining models from scratch.
Second, \zl{it introduces} a lightweight context-aware router that maps query contexts directly to expert combinations.
By decoupling routing from execution, \MyArch ensures contiguous memory access via homogeneous batching, thereby 
\lyy{avoiding the extra buffering overhead }
and reducing peak activation memory.
To further address system-level bottlenecks, we design an asynchronous CPU–GPU execution pipeline with hierarchical bucketing that masks preprocessing overhead and reduces padding-induced waste.
We summarize our main contributions as follows:
\begin{itemize}[leftmargin=*]
    \item We formulate query context-aware transformer slicing for in-database predictive query processing and propose QCATS, which uses query predicates and metadata statistics to specialize neural inference execution paths.
    \item We introduce a parameter decomposition strategy that partitions FFN layers into shared and semantic components, enabling efficient sparse inference from a fine-tuned dense Transformer without constructing separate expert networks from scratch.
    \item We design a lightweight query-level routing mechanism that maps query contexts to sparse expert combinations, shifting conditional computation from token-level dynamic routing to query-level static routing.
    \item We integrate QCATS into a database-oriented execution pipeline with routing-aware batching and asynchronous CPU–GPU execution.
    Experiments on four workloads and two Transformer backbones show significant latency and memory reductions while preserving prediction accuracy.
\end{itemize}

\noindent
The rest of the paper is organized as follows. 
\S\ref{sec:foundations} defines predictive queries and formulates query context-aware transformer slicing.
\S\ref{sec:method} presents \MyArch and
\S\ref{sec:sys_opt} introduces database-oriented execution optimizations. 
\S\ref{sec:experiments} reports the experimental evaluation. 
\S\ref{sec:related} reviews related work, and \S\ref{sec:conclusion} concludes the paper.

\section{Foundations}
\label{sec:foundations}
In this section, we establish the foundations required to reason about query-aware model slicing in database systems. 

\begin{figure*}[t]
    \begin{minipage}[t]{0.58\linewidth} 
        \centering
        \begin{subfigure}[t]{0.32\linewidth}
            \centering
            \vspace{0pt}
            \includegraphics[width=\linewidth]{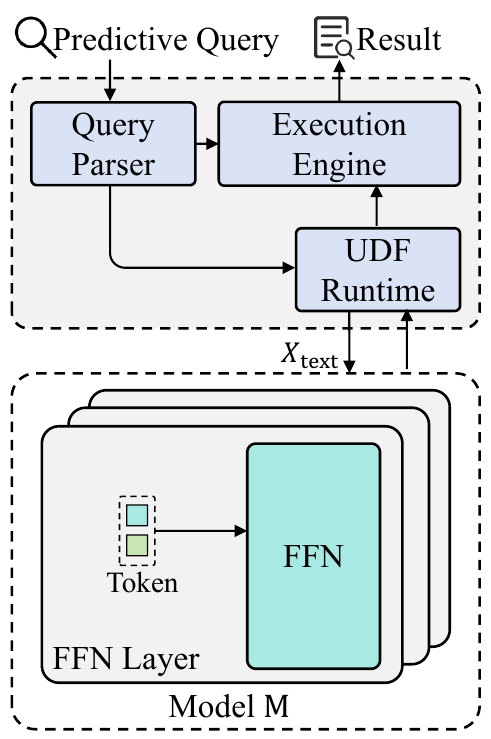}
            \caption{With Dense Model}
            \label{fig:2a}
        \end{subfigure}%
        \hfill
        \begin{subfigure}[t]{0.32\linewidth}
            \centering
            \vspace{0pt}
            \includegraphics[width=\linewidth]{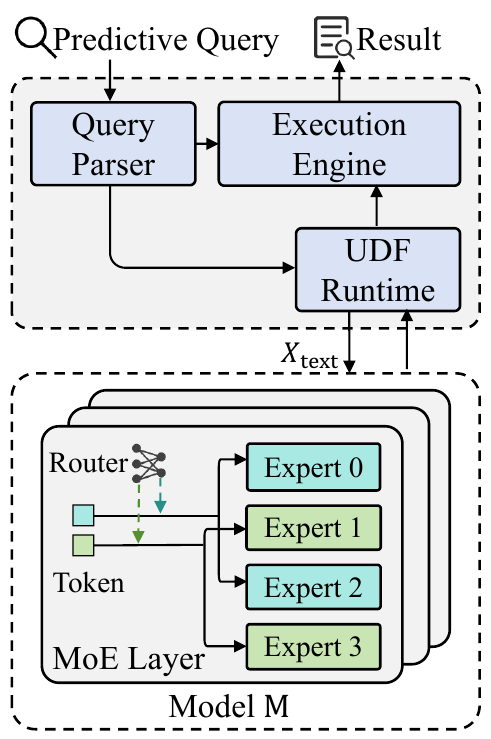}
            \caption{With MoE Model}
            \label{fig:foundation_left}
        \end{subfigure}%
        \hfill
        \begin{subfigure}[t]{0.32\linewidth}
            \centering
            \vspace{0pt}
            \includegraphics[width=\linewidth]{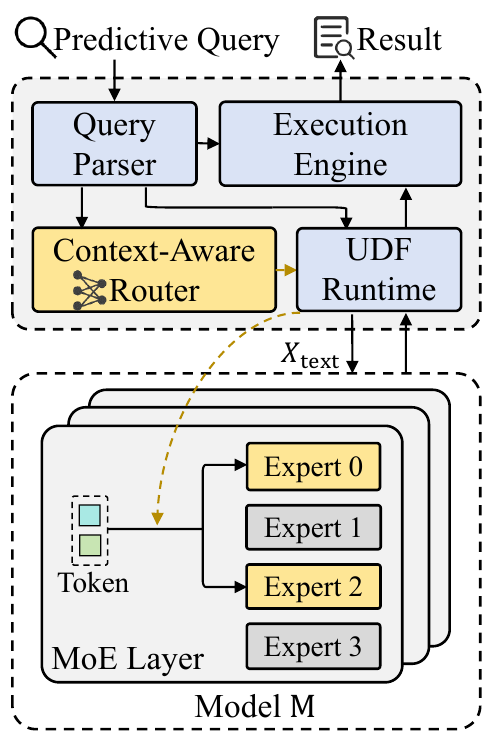}
            \caption{With \MyArch(Ours)}
            \label{fig:foundation_right}
        \end{subfigure}
        
        \caption{Predictive Query Processing in Database.}
        \label{fig:figure2_main}
    \end{minipage}%
    \hfill 
    \begin{minipage}[t]{0.41\linewidth}
        \centering
        \vspace{0pt}
        \includegraphics[width=\linewidth]{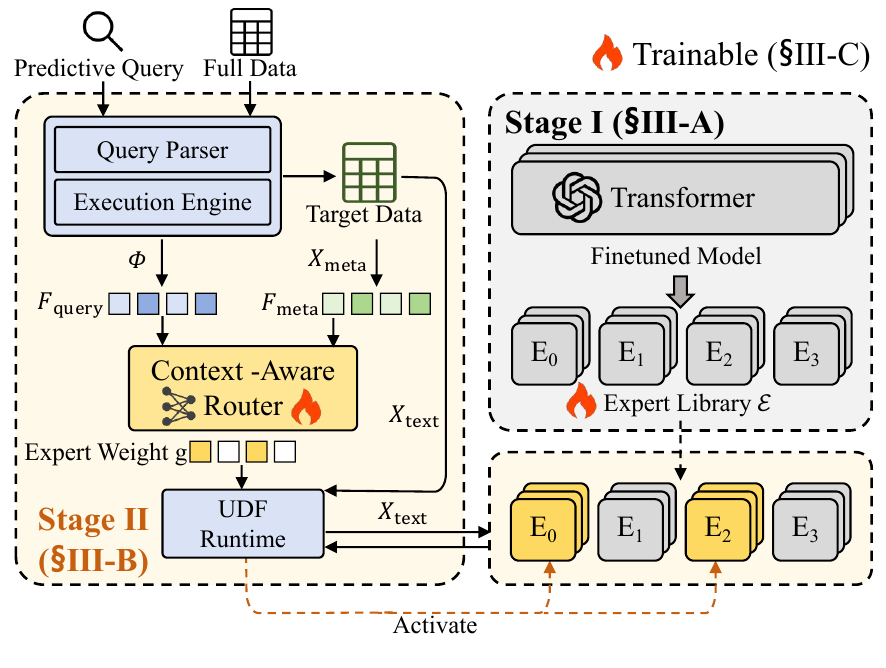}
        
        \caption{Overview of \MyArch.}
        \label{fig:overview}
    \end{minipage}
\end{figure*}
\subsection{Predictive Query}

We consider \textit{predictive queries} in analytical databases, which integrate relational filtering with model inference. 

Let a relation $\mathcal{T}$ consist of text attributes $\mathcal{X}_{\text{text}}$ and structured metadata $\mathcal{X}_{\text{meta}}$.
The metadata describes the context of each text record, such as product category, brand, or user profile in e-commerce scenarios.
Crucially, our framework is broadly applicable to any scenario where the model input is shaped by SQL evaluation over the metadata. 
Given relational constraints $\Phi$ over $\mathcal{X}_{\text{meta}}$, a predictive query $Q$ first obtains an intermediate subset $\mathcal{T}_{\Phi(\mathcal{X}_{\text{meta}})}$ through standard relational operators, including \texttt{WHERE}, \texttt{JOIN}, and \texttt{GROUP BY}, and then applies a Transformer-based model $\mathcal{M}$ to $\mathcal{X}{\text{text}}$.

Formally, we define the predictive query $Q$ as an extended projection over the metadata-constrained subset:
\begin{equation}
    \mathcal{T}_{\text{scored}} = \pi_{*, \mathcal{M}(\mathcal{X}_{\text{text}}) \to Y} \left( \mathcal{T}_{\Phi(\mathcal{X}_{\text{meta}})} \right)
\end{equation}
where $\mathcal{T}_{\Phi(\mathcal{X}_{\text{meta}})}$ denotes the intermediate subset materialized by the database engine under metadata constraints $\Phi$, and $\mathcal{M}$ evaluates $\mathcal{X}_{\text{text}}$ to append a prediction column $Y$.

We define \textbf{query context} $\mathcal{C}_Q$ as a composite representation:
\begin{equation}
\mathcal{C}_Q = \bigl(\,\Phi,\;\mathcal{S}(\mathcal{T}_\Phi)\,\bigr)
\end{equation}
where $\Phi$ denotes explicit relational constraints, and $\mathcal{S}(\mathcal{T}_\Phi)$ captures statistics over the post-constraint subset $\mathcal{T}_\Phi$.
The context $\mathcal{C}_Q$ specifies the semantic scope of model inference and can be materialized into dense features via feature engineering or lightweight encoders, as seen in \S\ref{sec:context}.
Note that existing UDF-based inference processes tuples independently and does not expose such relational context to the inference pipeline.

\subsection{Core Idea and Goal}
\label{sec:goal}

Transformer feed-forward networks (FFNs) dominate model parameters and inference cost~\cite{vaswani2017attention,fedus2022switch}.
Yet, their activations are highly sparse: prior studies show that retaining a modest subset of neurons can preserve most output magnitude and predictive performance~\cite{lilazy,liu2023deja,zhang2022moefication}.
This redundancy motivates sparse FFN execution.
Conditional computation mechanisms such as Mixture-of-Experts (MoE) exploit this redundancy by replacing dense FFNs with expert networks ${E_1,\dots,E_N}$ controlled by a routing function $G(\cdot)$.
Formally, for an input hidden state $\mathbf{x}$, the output is:
\begin{equation}
    \mathbf{y} = \sum_{i=1}^{N} G(\mathbf{x}) \cdot E_i(\mathbf{x}).
\end{equation}
Existing MoE architectures are primarily designed for large-scale pre-training and rely on \textit{token-level, layer-wise} routing, in which routing decisions are computed dynamically for individual tokens.
While effective for scaling model capacity, this fine-grained routing mechanism introduces non-trivial dynamic scheduling and dispatching overhead and fails to exploit the semantic coherence exposed by database query execution.

As illustrated in Fig.~\ref{fig:figure2_main}, our \textbf{key insight} is that predictive queries naturally enable \textit{query-level conditional execution}.
Different from token-level conditional computation in common MoE design (Fig.~\ref{fig:foundation_left}), which adapts execution paths independently for each input, query-level conditional execution (Fig.~\ref{fig:foundation_right}) leverages the semantic consistency of query-selected subsets and applies a unified execution configuration across all tuples in $\mathcal{T}_{\Phi}$.

Under this paradigm, the query context acts as a high-level routing control signal for model inference activation.
Routing signal can be computed before inference, enabling static execution paths, homogeneous batch construction, and tighter integration with relational execution pipelines.

\textbf{Goal.} 
\yy{
We focus on predictive queries that apply Transformer-based models independently to many tuples. 
Our objective is to reduce the accumulated inference cost over $\mathcal{T}_{\Phi}$ rather than improving autoregressive generation and serving-specific mechanisms of the Transformer itself.
}

We formalize our task as \textbf{Query Context-Aware Transformer Slicing} problem.
Let $\theta$ denote the parameter set of a Transformer model $\mathcal{M}$.
Given a predictive query $Q$ with constraints $\Phi$, our goal is to generate a sliced model $\mathcal{M}_{\Phi}$ that selects a sparse parameter subset $\theta_{\Phi}\subset\theta$ for efficient query execution.
The sliced model should satisfy two goals:
\begin{itemize}[leftmargin=*, labelindent=0pt]
    \item \textbf{Execution Efficiency (G1).} 
    The sliced model should reduce inference latency and peak memory footprint when integrated into the database query pipeline.

    \item \textbf{Accuracy Preservation (G2).} 
    The activated slice should preserve the predictive performance of the full model on the target subset $\mathcal{T}_{\Phi}$ within a small tolerance margin.
\end{itemize}

\section{\MyArch}
\label{sec:method}
We propose \textbf{Query Context-Aware Transformer Slicing (\MyArch)}, a framework that shifts model inference from dense execution to query-adaptive sparse execution. Unlike traditional MoE with token-level routing, \MyArch adopts query-level routing to pre-activate semantic parameter subsets, reducing computation and VRAM usage while preserving predictive accuracy.
\lyy{As shown in Fig.~\ref{fig:overview}, \MyArch consists of two stages: offline expert construction and query-time dynamic assembly.}
\yy{
Note that the offline construction is performed once while the online path performs only lightweight routing and sliced inference.}

\textit{Stage I: Offline Sparsity-Based Expert Construction.}

\zl{
\MyArch exploits FFN activation sparsity to extract sparse expert slices from a fine-tuned dense Transformer. 
These slices form an expert library $\mathcal{E}$ that supports conditional execution while preserving the model's semantic capability, as elaborated on \S\ref{sec:structural_prep}.
}

\textit{Stage II: Query-Time Context-Aware Inference.}
\zl{
At query time, \MyArch extracts relational constraints and metadata statistics as query context, maps them to expert weights through a lightweight router, and executes inference using the selected slices. The inference pipeline, as detailed in \S\ref{sec:context} consists of Query Processing, Query Context Featurization, Context-Aware Expert Routing, and Sliced Model Execution.
}
\subsection{Stage I: Expert Construction}
\label{sec:structural_prep}

\begin{figure}[t]
    \centering
    \includegraphics[width=\columnwidth]{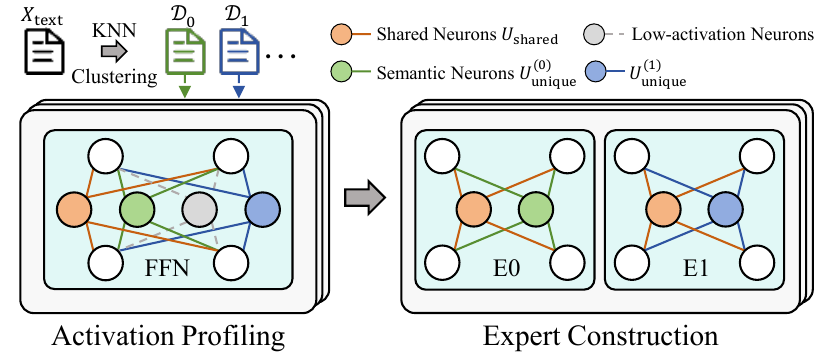}
    \caption{Illustration of Expert Construction.}
    \label{fig:expert_construction}
\end{figure}

\MyArch begins with a one-off structural reconfiguration step performed during system initialization to prepare the dense transformer model for query-level conditional execution. 
By analyzing activation patterns in FFNs, we decompose the original model into a library of physically isolated \emph{Expert Slices} as illustrated in Fig.~\ref{fig:expert_construction}.
This offline transformation converts monolithic FFN layers into modular computation units that can be selectively activated at query time.
The resulting Expert Library provides a stable execution substrate for query-level routing and selective parameter activation, enabling efficient, deterministic, and database-friendly sparse inference.

To construct specialized experts, \MyArch first identifies distinct \lyy{semantic regions from the text data $X_{\text{text}}$.}
Directly using all data samples introduces noise due to overlapping topics and ambiguous semantic boundaries, which can blur the specialization signals required for expert construction.
Instead, we \zl{choose} to extract prototypical \lyy{samples}.
Specifically, we employ a lightweight embedding model (FastText~\cite{mikolov2018advances}) to extract embeddings and apply K-Means clustering to partition the data into $N$ \lyy{semantic} groups ${\mathcal{D}_1, \dots, \mathcal{D}_N}$
\footnote{
\zl{
We set the sample size of each group to 500 as a trade-off between effectiveness and cost.
}
}.
For each group, we retain only the samples closest to the corresponding cluster centroid \zl{to reduce noise}. These representative \lyy{samples} provide stable and discriminative activation patterns for subsequent neuron analysis, ensuring that each expert slice is anchored to a well-separated semantic domain.

Using these samples, we profile the internal activation behavior of the fine-tuned dense model. 
We perform forward passes on data from each \lyy{semantic} group $\mathcal{D}_n$ and collect neuron activation statistics across FFN layers. 
This profiling step maps semantic variations in the input space to structured patterns in the \lyy{neuron activation } space, producing the activation signals required to identify different types of neurons.

Neurons in FFN layers exhibit heterogeneous functional roles~\cite{cai2024textit}: some encode general linguistic patterns that are broadly useful across domains, while others respond selectively to specific \lyy{semantic regions.}
To preserve both generalization capability and domain specialization, \MyArch explicitly separates these two types of neurons.
Concretely, we partition FFN neurons into a shared set $U_{\text{shared}}$, which captures universally useful computation, and multiple \lyy{semantic} sets $U_{\text{unique}}^{(n)}$, each corresponding to a distinct \lyy{semantic} group $\mathcal{D}_n$. 

First, to capture general knowledge shared across different domains, we aim to identify neurons that remain consistently active regardless of the specific \lyy{semantic} group.
To quantify the universal utility, we calculate General Importance Score $S_j^{\text{global}}$:
\begin{equation}
S_j^{\text{global}} = \frac{1}{N} \sum_{n=1}^{N} A_j^{(n)}
\end{equation}
where $A_j^{(n)}$ denotes the average activation magnitude of neuron $j$ in \lyy{semantic} group $n$, and $N$ is the total number of groups.
Neurons with the highest scores are then selected to form the shared set $U_{\text{shared}}$, ensuring the retention of fundamental model capabilities while providing a stable foundation for subsequent \lyy{semantic} specialization.

Subsequently, to capture \lyy{semantic} knowledge, we need to identify neurons that exhibit high specificity.
Such neurons are highly active in a specific \lyy{semantic} group but suppressed in other groups.
To quantify this contrastive property, we define \lyy{Semantic} Score $R_{j,n}$:
\begin{equation}
R_{j,n} = \frac{A_j^{(n)}}{\frac{1}{N-1}\sum_{m \neq n} A_j^{(m)} + \epsilon}
\end{equation}
where the denominator represents the background noise level from other groups.
By filtering for neurons with top-ranked $R_{j,n}$ values, we construct the \lyy{semantic} set $U_{\text{unique}}^{(n)}$ for each \lyy{semantic group, ensuring that each expert specializes in its corresponding group.}

Based on the selected neuron sets, we physically reconstruct the original dense parameter matrix $\mathbf{W}$ into $N$ isolated Expert Slices, namely the tensor set $\{ \mathbf{W}^{(1)}, \dots, \mathbf{W}^{(N)} \}$. \lyy{For standard Transformer FFN layers composed of up projection $\mathbf{W}_{up}$, gate projection $\mathbf{W}_{gate}$, and down projection $\mathbf{W}_{down}$, we apply neuron selection consistently across all three matrices to ensure
structural integrity.}
Specifically, the $n$-th expert slice $\mathbf{W}^{(n)}$ is composed of the union of the shared set and the corresponding \lyy{semantic} set, i.e., $U_{\text{shared}} \cup U_{\text{unique}}^{(n)}$. 
This design allows each expert slice to independently retain both general linguistic capabilities and \lyy{semantic} expertise.

\subsection{Stage II: Context-Aware Inference}
\label{sec:context}
At query time, \MyArch extracts the query context $\mathcal{C}_Q$ from each incoming query. 
By matching it against the $N$ pre-established semantic groups, the system routes the query to the corresponding expert for efficient inference.
Specifically, we represent $\mathcal{C}_Q$ as a dense context embedding $e_{\text{ctx}}$ through feature engineering and a lightweight encoder, then use it to drive expert routing. 
This process comprises the following four key steps:

\subsubsection{Query Processing}
\begin{figure*}[t]
    \begin{minipage}[t]{0.62\linewidth} 
        \centering
        \begin{subfigure}[t]{0.585\linewidth}
            \centering
            \vspace{0pt}
            \includegraphics[width=\linewidth]{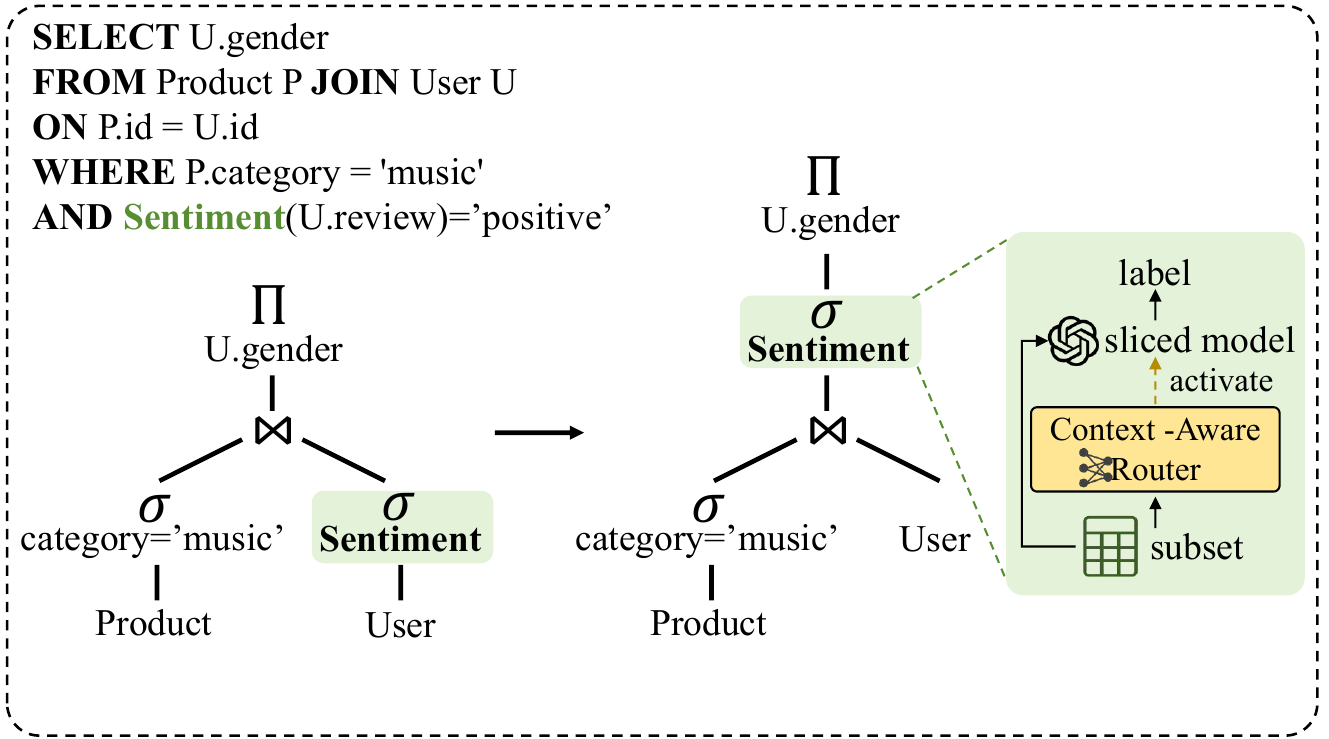} 
            \caption{Query Rewriting}
            \label{fig:query_rewrite}
        \end{subfigure}%
        \hfill
        \begin{subfigure}[t]{0.41\linewidth}
            \centering
            \vspace{0pt}
            \includegraphics[width=\linewidth]{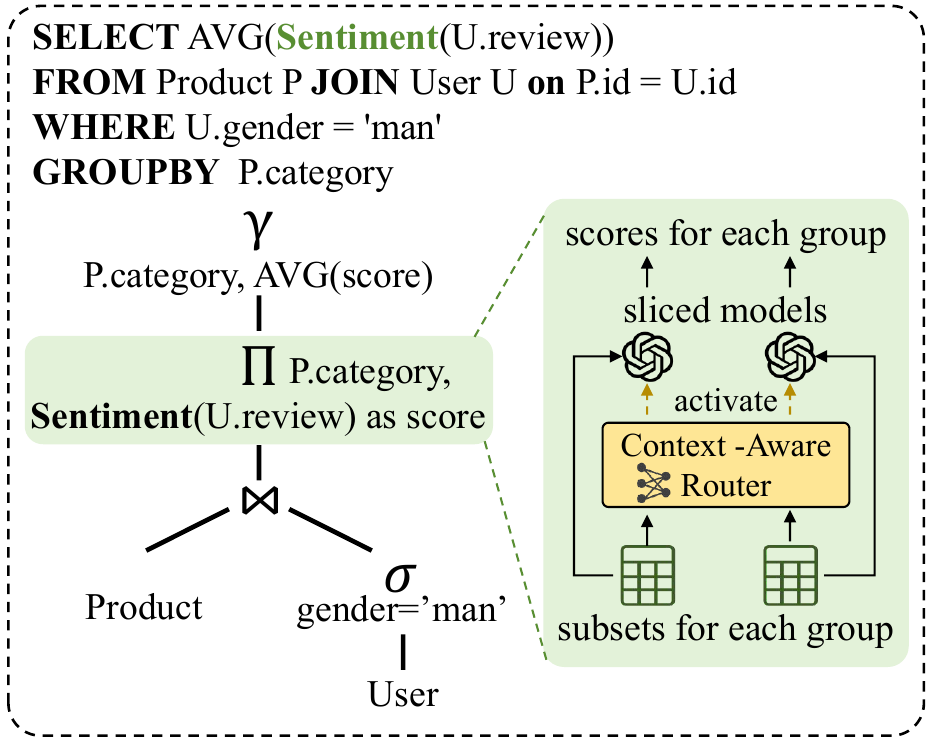} 
            \caption{Grouped Query Processing}
            \label{fig:query_group}
        \end{subfigure}
        
        \caption{Specialized Query Processing in QCATS.}
        \label{fig:main_figure_queries}
    \end{minipage}%
    \hfill 
    \begin{minipage}[t]{0.375\linewidth}
        \centering
        \vspace{0pt}
        \includegraphics[width=\linewidth]{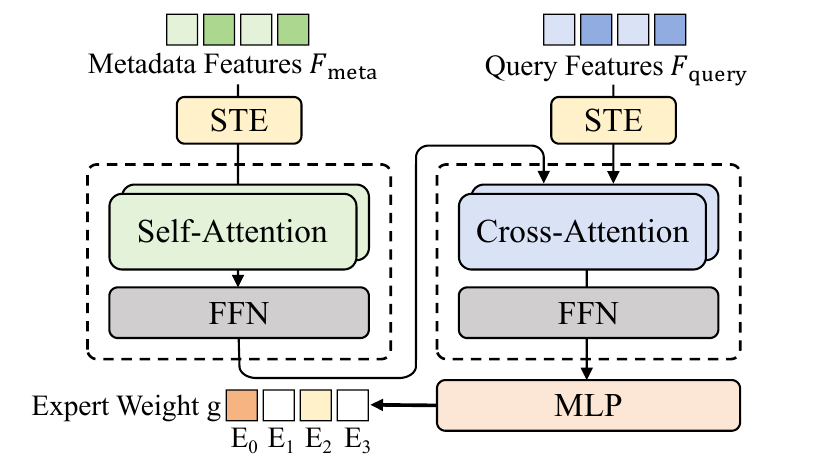} 
        \vspace{0cm}
        \caption{Architecture of Context-Aware Router.}
        \label{fig:encoder}
    \end{minipage}
\end{figure*}

Raw predictive queries often interleave lightweight relational operators (e.g., \texttt{JOIN}, \texttt{GROUP BY}) with computationally expensive neural UDFs. Because the data fed into the neural model is strictly pre-filtered by these SQL operators, \MyArch must first parse the query to identify these filtering boundaries. We formally extract these relational filtering signals as explicit constraints, denoted as $\Phi$. This query parsing and semantic-aware logical rewriting serve two main purposes: minimizing the data subset fed into the inference stage and providing highly selective semantic context $\Phi$ for downstream routing.

A major challenge in extracting a clean $\Phi$ arises when neural UDFs are embedded within predicates. 
In database query processing, the query parser translates an input SQL into a logical query plan represented as an Abstract Syntax Tree (AST), where relational predicates and UDF invocations coexist as distinct operator nodes.
A conventional optimizer may treat neural UDF predicates similarly to scalar predicates or lack sufficient cost information to delay them appropriately.
To prevent this, \MyArch introduces a \textit{deferred evaluation rule}. 
As shown in Fig.~\ref{fig:query_rewrite}, it rewrites the AST by moving the UDF predicate ($\sigma$) upward, past lightweight scalar constraints and \texttt{JOIN}s, delaying its execution. 
This deferral naturally isolates all preceding relational operators into the explicit constraints $\Phi$. 
Consequently, the UDF acts as a final semantic filter evaluated on a reduced subset, while $\Phi$ provides a precise semantic context for routing.
Crucially, features are always extracted from \zl{the} subset that the UDF actually processes, so the routing input remains consistent with the true inference domain regardless of how the rewrite reshapes the data.

For queries involving logical partitioning (e.g., \texttt{GROUP BY}), \MyArch abandons standard tuple-at-a-time inference (Fig.~\ref{fig:query_group}). Instead, the execution engine physically partitions intermediate tuples into dense semantic subsets prior to model inference. The context-aware router then treats each pre-aggregated subgroup as an independent batch, activating distinct sliced models. 
To balance context-awareness with overhead, QCATS ranks the pre-aggregated groups by their cardinalities in the query-filtered subset, assigns dedicated expert slices only to the few dominant groups, and merges the remaining tail groups into a shared routing path.

Ultimately, this parsing and rewriting phase outputs a refined logical plan alongside the clearly isolated explicit constraints $\Phi$. These elements define the exact semantic boundaries of the target data, providing a precise, noise-free foundation for the subsequent featurization phase.

\subsubsection{Query Context Featurization}
Discrete SQL constraints and structured metadata are semantically heterogeneous and cannot be directly processed by a neural router.
To bridge the gap between relational structures and continuous embeddings, \MyArch performs a two-phase preparation procedure:
\textit{Feature Selection}, which identifies informative metadata attributes, and \textit{Context Representation}, which transforms them into unified vector representations suitable for downstream routing.

\uline{\textit{One-time Feature Selection.}}
Database tables often contain redundant or weakly informative metadata that contributes little to semantic understanding. 
Feeding all columns to the router introduces noise and computational overhead. 
To address this issue, we design a dual-view scoring mechanism 
that identifies a compact subset of highly discriminative attributes. 
Importantly, this process is executed as a one-off offline task and incurs no overhead during query-time inference.

To quantify the global discriminative power of each column, we train a lightweight classifier using metadata attributes as input features.
Specifically, we measure the global relevance score of each column using Gini importance derived from a Random Forest model~\cite{breiman2001random} trained to predict the context groups $\{\mathcal{D}_n\}$ built in \S\ref{sec:structural_prep}.

In ambiguity zones, textual features alone are insufficient.
We must identify columns that distinguish samples where text fails to do so. 

To this end, we construct a KNN~\cite{cover1967nearest} index over the training set of the target downstream task using text embeddings, and retrieve each sample's $k$ nearest neighbors. Here, the downstream task labels are the ground-truth labels provided in that training set (e.g., sentiment labels in a sentiment analysis dataset of product reviews).
A sample pair is marked as a \textit{conflict pair} if the two samples are close in embedding space yet carry different labels from this training set, indicating that textual similarity alone cannot reliably discriminate them.
We then compute the average value discrepancy of each metadata column across all identified conflict pairs, and adopt this discrepancy as the local disambiguation score, prioritizing columns that differ systematically across such pairs and therefore may help resolve ambiguities left by text alone.
Finally, we select the Top-$M$ columns based on a weighted average of these scores, defining them as the effective \textbf{context columns}.

\uline{\textit{Schema-Agnostic Featurization.}}
To ensure the router generalizes across diverse database tables, we require a standardized input format that abstracts away schema differences.
We introduce an input featurization mechanism that maps diverse columns into a unified embedding space.
Numerical values are processed via discretization-based encoding,
while categorical fields utilize frequency-based encoding.
The function $\mathcal{F}(\cdot)$ generates two dense feature sets that enable the model to infer explicit intent against global data distributions without relying on fixed position rules:

\noindent
\textbf{(a) Implicit Metadata Features} $\mathbf{F}_{\text{meta}}$:
This feature set captures the statistical characteristics of the selected context columns under the current dataset distribution. 
It encodes both column semantics and distributional profiles as:
\begin{equation}
    \mathbf{F}_{\text{meta}} = \{ \mathbf{e}_{\text{col}}^{(m)} , \mathbf{e}_{\text{stat}}^{(m)} \}_{m=1}^{M} 
\end{equation}
where $M$ denotes the number of selected context columns.
For the $m$-th column, $\mathbf{e}_{\text{col}}^{(m)}$ represents the column name embedding, and $\mathbf{e}_{\text{stat}}^{(m)}$ encodes its statistical distribution properties.

\noindent
\textbf{(b) Explicit Predicate Features} $\mathbf{F}_{\text{query}}$:
This feature set captures the user's specific filtering intent expressed in SQL predicates. To minimize noise from irrelevant attributes, we retain only predicates involving the selected context columns.
If the query contains $J$ such predicates, the representation is constructed as:
\begin{equation}
    \mathbf{F}_{\text{query}} = \{ \mathbf{e}_{\text{col}}^{(j)}, \mathbf{e}_{\text{val}}^{(j)} \}_{j=1}^{J}
\end{equation}
where $J$ is the number of valid predicates.
For the $j$-th predicate, $\mathbf{e}_{\text{col}}^{(j)}$ and $\mathbf{e}_{\text{val}}^{(j)}$ denote the embeddings of the target column and filtering value, respectively. 
Note that if there exist no relevant predicates, this feature set is empty, i.e., $J=0$.


\subsubsection{Context-Aware Expert Routing}
\label{sec:context_awareness}
To transform the query context features into specific expert configurations, we design a customized Context-Aware Router.
Unlike traditional MoE architectures that perform token-level, layer-wise routing inside the model, our router operates as a decoupled and lightweight look-ahead module that determines expert activation prior to inference.
Our design is inspired by hybrid attention architectures~\cite{li2023alece}, but is specifically adapted to effectively process the non-sequential and structured nature of database content, as depicted in Fig.~\ref{fig:encoder}.

Since database columns are permutation-invariant, standard positional encodings are unsuitable as they impose artificial order.
To preserve the native structure of tabular inputs, we discard absolute positioning and introduce Structural Type Embedding (STE),
enabling the router to capture relational semantics without spurious sequential dependencies.

Building on this structured input, we employ a specialized Transformer model interaction to distill semantic context. 
First, to capture global correlations among columns (e.g., the joint distribution between \texttt{Brand} and \texttt{Price}), we utilize a 2-layer Transformer encoder to process the implicit metadata features $\mathbf{F}_{\text{meta}}$, generating a contextually consistent global representation $\mathbf{H}_{\text{meta}}$. 

Next, to filter out redundant signals and focus solely on the query-specific intent, we modify a 2-layer Transformer decoder by repurposing the Multi-Head Cross-Attention (MHCA) mechanism. 
We initialize the query state $\mathbf{C}^{(0)}$ using the explicit predicate features $\mathbf{F}_{\text{query}}$, or employ a learnable global intent embedding $\mathbf{e}_{\text{global}}$ to autonomously discern salient metadata when predicates are absent.
With the encoded global metadata $\mathbf{H}_{\text{meta}}$ serving as the \textbf{key} and \textbf{value},
the decoder dynamically attends to metadata attributes that are most relevant to the current query context:

\begin{equation}
\mathbf{C}^{(l)} = \text{LayerNorm}(\mathbf{C}^{(l-1)} + \text{MHCA}(\underbrace{\mathbf{C}^{(l-1)}}_{\text{Query}}, \underbrace{\mathbf{H}_{\text{meta}}}_{\text{Key}}, \underbrace{\mathbf{H}_{\text{meta}}}_{\text{Value}}))
\end{equation}
where $\mathbf{C}^{(l)}$ denotes the context representation at decoder layer $l$.

The final stage transforms the variable-length semantic context into a discrete, efficient execution plan. 
We first apply Masked Mean Pooling on the decoder output to produce a fixed-length context embedding $\mathbf{e}_{\text{ctx}}$.
To explicitly bound inference cost, we then employ a Top-$K$ Gating Network, which projects $\mathbf{e}_{\text{ctx}}$ through an MLP and activates only the Top-$K$ most relevant expert slices:
\begin{equation}
\mathbf{g} = \text{Top-K}(\text{Softmax}(\text{MLP}(\mathbf{e}_{\text{ctx}})))
\end{equation}
where $\mathbf{g} \in \mathbb{R}^N$ (with only $K$ non-zero entries) represents the sparse gating vector containing the routing weights for the selected experts. 
The hyperparameter $K$ explicitly governs the trade-off between model capacity and execution efficiency: a smaller $K$ prioritizes computational sparsity, while a larger $K$ preserves richer semantic nuances.
This design effectively translates semantic understanding into a deterministic, hardware-friendly execution path, eliminating redundant parameter computations.

\subsubsection{Sliced Model Execution}

In the inference phase, once the context-aware router generates routing decisions, the system directly retrieves the activated Top-$K$ expert tensors $\{\mathbf{W}^{(k)} \mid k \in \mathcal{E}_{\text{active}}\}$ from the Expert Library. 
The output $\mathbf{y}$ of the sliced Transformer FFN layer is computed as:
\begin{equation}
\mathbf{y} = \sum_{k \in \mathcal{E}_{\text{active}}} \mathbf{g}_k \cdot E_k(\mathbf{x})
\end{equation}
where $\mathbf{x}$ denotes the input hidden state and $\mathbf{g}_k$ is the expert weight.

As mentioned previously, unlike token-level MoE execution, our sliced model execution is performed at query granularity and uses pre-constructed expert slices, avoiding dynamic control flow and runtime expert construction.
To be specific, the design yields two key efficiency benefits.
First, QCATS reduces computation associated with inactive expert slices by restricting FFN matrix multiplications to the selected parameter subset.
Second, peak VRAM reduction is achieved because expert slices have substantially smaller dimensionality than the original dense FFN. 
Consequently, the size of intermediate activations decreases proportionally, eliminating the need to allocate large GPU memory buffers for the full-width network and thereby enabling memory-efficient deployment.

\textbf{Discussion.}
The practicality of \MyArch is demonstrated through its robust design and seamless system integration. 
Its architecture provides adaptive semantic coverage: queries with strong semantic signals are routed to dominant specialized experts, while weakly constrained queries benefit from weighted multi-expert collaboration. 
The shared neuron set $U_{\text{shared}}$ preserves fundamental linguistic capabilities, ensuring a stable performance floor across diverse query contexts.

\yy{
It is worth noting that QCATS complements general-purpose model-serving platforms and low-level execution optimizations. While vLLM~\cite{kwon2023efficient} and SGLang~\cite{zheng2024sglang} optimize autoregressive serving through KV-cache management and continuous batching, QCATS uses query context to reduce short-output inference costs across database tuples. 
QCATS also supports attention acceleration methods, such as FlashAttention~\cite{dao2022flashattention}, FastMoE~\cite{he2021fastmoe}, and quantization~\cite{frantar2023optq,lin2024awq,xiao2023smoothquant}. 
Decoupling the expert library from the router enables efficient execution under stable data distributions.
As data and workloads gradually evolve, incremental fine-tuning of the lightweight router can support long-term deployment in dynamic OLAP scenarios without rebuilding the expert library.
}

\subsection{Decoupled Training Strategy}
\label{train}

To operationalize \MyArch, we must optimize two distinct components: the Context-Aware Router $\psi$ and the Expert Library $\mathcal{E}$.
Since the router is initialized from scratch, while experts are physically derived from a dense backbone, naive joint optimization \zl{may lead} to unstable training dynamics. 
Specifically, the router requires specialized experts to learn discriminative routing policies, whereas expert specialization depends on stable and consistent routing decisions. 

To address this \zl{problem}, we adopt a decoupled two-phase training strategy. 
\textit{Phase I} establishes a reliable routing prior through teacher-guided semantic alignment, while \textit{Phase II} leverages the fixed routing policy to drive physical expert specialization.
This separation stabilizes optimization and enables effective coordination between routing and expert learning.

\subsubsection{Phase I: Router Semantic Alignment via Activation Distillation}

In this phase, we train Context-Aware Router by constructing supervised routing targets derived from the internal activation behavior of the fine-tuned dense model. 
While Section~\ref{sec:structural_prep} determines the physical composition of expert slices, the semantic correspondence between query contexts and experts remains undefined. 
We bridge this gap by treating the dense model as a teacher and using its FFN activations as oracle signals to align the expert with the query context.

\zl{
Given a query-selected subset $\mathcal{D}_q$, we run the dense model and aggregate activation magnitudes over the unique neurons of each expert:
}
\begin{equation}
s_{\text{raw}}^{(n)} = \frac{1}{|\mathcal{D}_q|} \sum_{x \in \mathcal{D}_q} \sum_{j \in U_{\text{unique}}^{(n)}} |h_j(x)|
\end{equation}
\zl{
where $h_j(x)$ is the post-activation output of neuron $j$. 
Note that shared neurons are excluded because they are active across most inputs and provide weak routing signals.
}

Even when focusing on unique neurons, disparities in absolute activation magnitudes persist across different experts.
To cope with this issue, we propose a calibration strategy that acts as a weighted balance between the relative improvement and the raw activation magnitude:
\begin{equation}
s_{\text{cal}}^{(n)} = \frac{s_{\text{raw}}^{(n)} - \mu_n}{\mu_n + \epsilon} + {\eta} \cdot s_{\text{raw}}^{(n)}
\end{equation}
where $\mu_n$ denotes the average activation magnitude of expert $n$ across the training corpus 
and $\eta$ is a coefficient controlling the weight of the original signal. 
This combination preserves the importance of strong raw signals while emphasizing context-specific spikes through the relative term.

\begin{equation}
\mathbf{g}_n^* = 
\begin{cases} 
\text{Normalize}(\text{Softmax}(\mathbf{s}_{\text{cal}} / \tau))_n & \text{if } n \in \mathcal{E}^* \\
0 & \text{otherwise}
\end{cases}
\end{equation}
where $\mathbf{g}_n^*$ denotes the ground truth routing weight for the $n$-th expert, $\tau$ is a temperature hyperparameter controlling the sharpness of the probability distribution. This formulation ensures that the ground truth labels not only identify the most suitable experts $\mathcal{E}^*$ but also accurately reflect their relative importance as determined by the teacher model. 

\subsubsection{Phase II: Context-Driven Joint Expert Training}

In the second phase, the context-aware router $\psi$ is frozen and serves as a fixed coordinator that determines data-to-expert assignment. 
The optimization objective shifts to the physical differentiation and joint training of the expert pool $\mathcal{E}$.
By decoupling router learning from expert optimization, we establish a stable training environment in which each expert is forced to specialize in the semantic subspaces identified during Phase I.

\paragraph{Router-Guided Physical Specialization}
To transform the dense FFN layers into a collection of specialized experts, we decompose the original model into structurally isolated expert modules $\{E_1, \dots, E_N\}$.
During forward propagation, expert execution strictly follows the routing decisions produced by the frozen router.
Only experts selected by the active index set $\mathcal{E}_{\text{active}}$ participate in computation and parameter updates:
\begin{equation}
\mathbf{y} = \sum_{k \in \mathcal{E}_{\text{active}}} \mathbf{g}_k \cdot E_k(\mathbf{x}; \theta_k)
\end{equation}
where $\mathbf{x}$ and $\mathbf{y}$ denote the input and output hidden states, respectively, and $\mathbf{g}_k$ represents the fixed routing weight assigned to expert $E_k$. 
Under this selective activation scheme, gradients are restricted to the parameters $\theta_k$ of active experts, enforcing strong functional separation and driving each expert to adapt exclusively to its designated context.

Recent studies on MoE refactoring~\cite{cai2024textit} show that routing patterns exhibit strong correlation across consecutive Transformer layers, rendering independent layer-wise routing largely redundant. 
Following this observation, we adopt a global routing consistency strategy, where the single routing vector $\mathbf{g}$ generated by $\psi$ is broadcast to all FFN layers. 
This constructs vertically aligned expert pathways that preserve semantic coherence throughout the network depth and enable consistent processing of specialized representations.

\paragraph{Distillation-Enhanced Expert Specialization (Optional)}
Directly transitioning from dense to sparse execution may introduce temporary performance degradation. 
To accelerate expert adaptation and preserve the teacher model's representational capacity, we incorporate the original dense model as a persistent reference and apply dual-granularity distillation~\cite{zuo2022moebert}.

Specifically, we introduce two auxiliary objectives:
\begin{enumerate}[leftmargin=*]
    \item \textbf{Hidden State Alignment.} We minimize the mean squared error between student ($S$) and teacher ($T$) hidden states across all Transformer layers. An adaptive normalization function $\mathcal{N}(\cdot)$ is applied to account for architectural differences (e.g., Pre-Norm vs. Post-Norm) and improve gradient stability:
    \begin{equation}
    \mathcal{L}_{\text{hid}} = \sum_{l=1}^{L} \text{MSE}(\mathcal{N}(\mathbf{H}_S^{(l)}), \mathcal{N}(\mathbf{H}_T^{(l)}))
    \end{equation}
    where $L$ denotes the total number of layers, and $\mathbf{H}_S^{(l)}$ and $\mathbf{H}_T^{(l)}$ represent the corresponding hidden states.
    
    \item \textbf{Logit Alignment.} We apply symmetric KL divergence to align output probability distributions, encouraging the sparse experts to preserve the predictive behavior of the dense teacher.
\end{enumerate}

\noindent
The overall optimization objective is defined as:
\begin{equation}
\mathcal{L}_{\text{total}} = \mathcal{L}_{\text{task}} + \alpha \left(\mathcal{L}_{\text{hid}} + \mathcal{L}_{\text{logit}} \right)
\end{equation}
where $\mathcal{L}_{\text{task}}$ denotes the primary task loss (e.g., the task in the predictive query), and $\alpha$ controls the strength of distillation regularization. 
This auxiliary supervision stabilizes sparse training and enables QCATS to retain dense-model accuracy under constrained execution budgets.

\section{System-level Optimization}
\label{sec:sys_opt}

In large-scale query scenarios, batching data from multiple queries is standard practice to maintain high GPU utilization. 
However, the diverse query contexts within mixed batches trigger divergent expert activation paths, which may fragment GPU parallelism and thereby degrade hardware efficiency.
Simultaneously, database text data typically exhibits a long-tailed length distribution, which makes static batching inefficient due to excessive zero padding. 
Furthermore, as \MyArch significantly boosts GPU inference efficiency, 
traditional synchronous execution pipelines could become increasingly CPU-bound, causing preprocessing latency to emerge as a major bottleneck for overall system performance.

\begin{figure}[t]
    \centering
    \includegraphics[width=\linewidth]{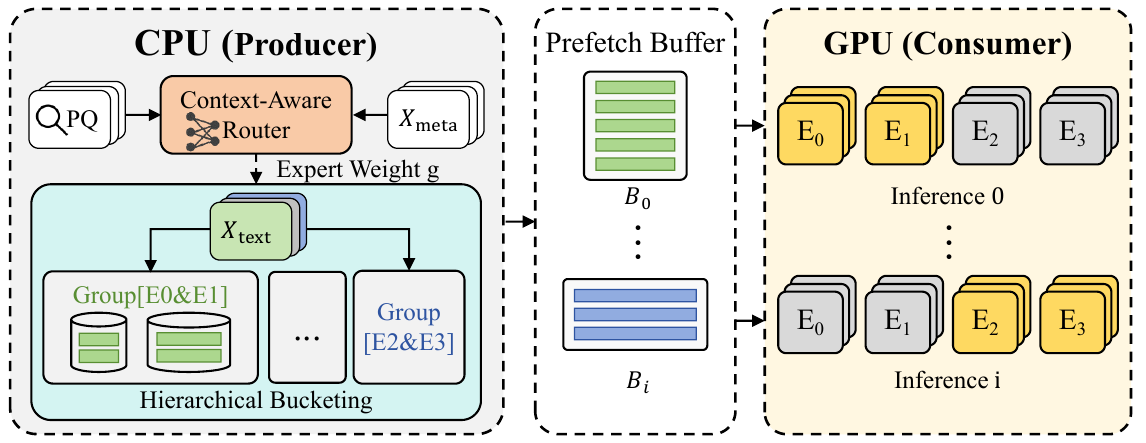}
    \caption{\MyArch Processing with Pipeline.}
    \label{fig:pipeline}
\end{figure}

To address the aforementioned inefficiencies, we propose a system-level optimization framework that adopts an asynchronous producer-consumer pipeline architecture.
As illustrated in Fig.~\ref{fig:pipeline}, we decouple the inference lifecycle into two separate modules:
The CPU acts as the producer, responsible for preparing routing-aware batched inputs,
while the GPU serves as the consumer, dedicated to performing compute-intensive sparse inference computations. 
A thread-safe prefetch buffer connects the two stages, allowing data preparation and model execution to proceed concurrently and overlap in time.

The core of the producer module, as shown in the left part in Fig.~\ref{fig:pipeline}, is \textbf{hierarchical bucketing}.
It reorganizes query data through a two-level process to jointly optimize execution-path consistency and hardware efficiency. 
The first level focuses on aligning model execution behavior by grouping inputs according to routing decisions, while the second level targets memory and computation efficiency by minimizing padding overhead through length-aware batching.

At the first level, QCATS leverages the context-aware router to precompute expert assignments before batch construction. 
Text data are grouped based on their routing outcomes, such that all texts within the same group activate an identical group of experts during inference. 
This routing-aware grouping ensures execution-path homogeneity within each batch and eliminates control-flow divergence on GPUs.

At the second level, text data within each routing group is further reorganized based on its length. 
Text entries with similar lengths are placed into the same bucket to reduce padding overhead. 
To maintain stable GPU utilization across varying input sizes, the system dynamically adjusts batch sizes according to the representative sequence length of each bucket and the available GPU resource budget. 
Subsequently, the bucketed text entries are tokenized and collated into batches, which are then enqueued into a prefetch buffer for immediate processing by the consumer module in the GPU.

In summary, the two-level reorganization enables QCATS to construct routing-consistent and length-homogeneous batches that maximize sparse execution efficiency while minimizing wasted computation caused by padding. 
Combined with the asynchronous producer–consumer pipeline, hierarchical bucketing effectively overlaps CPU-side preprocessing with GPU inference, eliminates pipeline stalls, and ensures that system throughput is primarily bounded by model execution performance rather than data preparation overhead.

\section{Experiments}
\label{sec:experiments}

In this section, we conduct extensive experiments to evaluate \MyArch against the design objectives mentioned in \S\ref{sec:goal}.
The evaluation mainly focuses on the following Research Questions (RQs):

\begin{itemize}[leftmargin=*]
     \item \textbf{RQ1 (System Efficiency):} 
     For \textbf{G1}, how much improvement does \MyArch achieve in terms of end-to-end latency and peak GPU memory footprint?
    \item \textbf{RQ2 (System Effectiveness):} 
    For \textbf{G2}, can \MyArch reproduce the predictive accuracy of the original fine-tuned dense baselines regarding the reduction in activated parameters?
    \item \textbf{RQ3 (Ablation \& Breakdown):} 
    What is the contribution of each component to the goals? Specifically, how does the latency breakdown demonstrate the effectiveness of our proposed optimization strategies? Meanwhile, what is the contribution of each system component to accuracy preservation?
    
    \item \textbf{RQ4 (Sensitivity \& Robustness):} 
    How sensitive is the model performance to key hyperparameters, and how is the neuron activation pattern aligned with the query context?
    In addition, does \MyArch maintain robustness when transferring to similar domains for data analysis? 
\end{itemize}

\subsection{Experimental Setup}
\label{subsec:setup}

\subsubsection{Datasets and Tasks}

\zl{
To evaluate \MyArch across diverse domains, we use four real-world datasets summarized in Table~\ref{tab:datasets}.
}
\begin{itemize}[leftmargin=*]
    \item \textbf{Amazon Review (E-commerce)}~\cite{hou2024bridging} contains user reviews and item metadata with 17 structured columns. The task is to predict product ratings from user reviews.
    \item \textbf{Kickstarter (Crowdfunding Finance)}~\cite{kantharia_kickstarter_v2} tracks crowdfunding campaigns and includes 7 structured metadata columns. The task is to predict campaign funding status from the project name and blurb.
    \item \textbf{OSHA (Industrial Safety)}~\cite{ou2025building} records industrial accidents with 12 structured metadata columns. The task is to classify injury severity from accident abstracts.
    \item \textbf{CVE (Cybersecurity)}~\cite{manzoni_vulnerability_mgmt} describes software vulnerabilities and contains 11 structured metadata columns. The task is to predict CVSS severity from vulnerability descriptions.
\end{itemize}

\begin{table}[t]
    \centering
    \caption{Summary of dataset statistics and task definitions.}
    \label{tab:datasets}
    \resizebox{\columnwidth}{!}{
    \begin{tabular}{l c c}
    \toprule
    \textbf{Dataset} & \textbf{Total Size} & \textbf{Context Columns} \\
    \midrule
    Amazon & 1,908,006 & [main\_category, sub\_category, price]  \\
    Kickstarter & 202,835 & [main\_category, sub\_category, goal]  \\
    OSHA & 42,645 & [occupation, accident\_type, age] \\
    CVE & 90,847 & [cwe, privileges\_required, product\_count]  \\

    \bottomrule
    \end{tabular}
    }
\end{table}




\subsubsection{Query Workload Generation}
\yy{
Public datasets rarely include query logs combining text inputs, structured metadata, and predictive UDFs, so we randomly generate 50 SQL templates with one to three predicates each for the datasets from the aforementioned widely-used benchmarks.
Context-column predicates use categorical equality or numerical ranges, combined only with \texttt{AND}; context columns may also appear in \texttt{GROUP BY} clauses. 
Non-context predicates are randomly sampled from \texttt{LIKE}, \texttt{IN}, equality, and range conditions, optionally connected by \texttt{OR}. 
Templates may include valid joins sampled from predefined joinable table pairs. 
These templates represent common analytical prediction queries involving category/range filtering, metadata joins, and grouped prediction aggregation. 
}


\subsubsection{Baselines and Implementation}
\zl{
To validate generality and effectiveness, we evaluate \MyArch on BERT and Qwen architectures, and compare three groups of methods:
}

\begin{enumerate}[leftmargin=*]
    \item \textbf{Dense Baselines:}
    \zl{
    Original unpruned BERT-base~\cite{devlin2019bert} and Qwen-base~\cite{yang2025qwen3} models.
    }

    \item \textbf{MoE Baselines:}
    \zl{
    MoEfication~\cite{zhang2022moefication}, which partitions FFN neurons into logical experts without changing weights, and MoEBERT~\cite{zuo2022moebert}, which converts a fine-tuned model into a physical MoE through adaptation and distillation.
    }

    \item \textbf{\MyArch (Ours):}
    \zl{
    \MyArch applied to BERT and Qwen, denoted as \textbf{\MyArch(BERT)} and \textbf{\MyArch(Qwen)}.
    }
\end{enumerate}

\zl{
All sparse methods activate 25\% of FFN parameters during inference, while dense baselines use all parameters. 
Since no standard MoEfication implementation exists for Qwen, we compare Qwen-base directly with \MyArch(Qwen).
}
\yy{
\MyArch is implemented in Apache Spark~\cite{apache_spark} 4.0.1 with
PyTorch 2.5 and CUDA 11.8, using BERT-base (110M parameters) and Qwen-0.6B
on an Intel Xeon Gold 5218R CPU and one NVIDIA A5000 GPU (24GB).
}

Note that we use compact models to reflect the resource constraints of predictive database systems and to isolate the benefits of query-aware sparsification.
Unless otherwise stated, \MyArch uses the N4K1 configuration ($N=4$, $K=1$), with the hidden dimension split between shared and unique neurons. 
The router uses the top $M=3$ attributes, a 512-hidden Transformer with 4 heads, and a 3-layer 256-dimensional MLP. 
During training, Phase I optimizes the router with $lr=5e-4$, $\eta=0.02$, and $ \tau=1.0$, while Phase II fine-tunes the experts with $\alpha=5.0$, and learning rates of $3e-5$ for BERT and $1e-5$ for Qwen, respectively.
Additional hyperparameters are reported in the appendix. 

\yy{\noindent\textbf{Offline Construction Cost.}
With 100k training examples, activation profiling, clustering, and expert-slice construction take a few minutes. 
The expert fine-tuning dominates offline cost, taking approximately 30 minutes for BERT and 2 hours for Qwen, or 1.5 and 6 hours with distillation, respectively.
}

\begin{table*}[t!]
    \centering
    \caption{End-to-end execution time (s) comparison. }
    \label{tab:e2e}
    \resizebox{\textwidth}{!}{
    \begin{tabular}{l | ccccc | ccc}
    \toprule
    \multirow{2.5}{*}{\textbf{Dataset}} & \multicolumn{5}{c}{\textbf{BERT-base Backbone}} & \multicolumn{3}{c}{\textbf{Qwen-base Backbone}} \\
    \cmidrule(lr){2-6} \cmidrule(lr){7-9}
& BERT-base & MoEfication & MoEBERT & \MyArch w/o pipeline & \textbf{\MyArch} & Qwen-base & \MyArch w/o pipeline & \textbf{\MyArch} \\
\midrule
    
    \textbf{Amazon} 
    & 279.05 & 299.32 (0.93$\times$) & 236.16 (1.18$\times$) & 164.02 (1.70$\times$) & \textbf{98.69 (2.83$\times$)}
    & 1499.22 & 517.49 (2.90$\times$) & \textbf{463.30 (3.24$\times$)} \\
    
    \textbf{Kickstarter} 
    & 352.94 & 376.14 (0.94$\times$) & 295.33 (1.20$\times$) & 172.89 (2.04$\times$) & \textbf{129.13 (2.73$\times$)}
    & 1739.19 & 874.38 (1.99$\times$) & \textbf{780.19 (2.23$\times$)} \\
    
    \textbf{OSHA} 
    & 785.40 & 847.95 (0.93$\times$) & 636.40 (1.23$\times$) & 307.46 (2.55$\times$) & \textbf{230.08 (3.41$\times$)}
    & 5241.14 & 2231.12 (2.35$\times$) & \textbf{1981.82 (2.64$\times$)} \\
    
    \textbf{CVE} 
    & 798.29 & 862.21 (0.93$\times$) & 653.82 (1.22$\times$) & 231.51 (3.45$\times$) & \textbf{180.51 (4.42$\times$)}
    & 4521.26 & 1564.17 (2.89$\times$) & \textbf{1418.08 (3.19$\times$)} \\
    
    \bottomrule
    \end{tabular}
}
\end{table*}
\subsubsection{Evaluation Settings and Metrics}
\label{sec:experimental}

\zl{
We evaluate efficiency by executing approximately 600 SQL queries per dataset, derived from the 50 query templates, retrieving 1 million rows in total. 
We report end-to-end latency and peak GPU memory usage, with the average batch size aligned across methods. 
We evaluate effectiveness using accuracy and weighted F1-score on the aggregated test set formed by merging query results from all workloads.
}

\subsection{System Efficiency (RQ1)}

To answer \textbf{RQ1}, we assess the system-level efficiency in end-to-end response time and peak memory.

\subsubsection{End-to-End Latency}
\label{sec:e2e}

Under the efficiency evaluation setting (\S\ref{sec:experimental}) on the 1-million-row workload, Table~\ref{tab:e2e} reports the end-to-end latency of all methods. All baselines and \textit{\MyArch w/o pipeline} use serial execution, whereas \textit{\MyArch} adopts the proposed pipelined architecture. The results show that \MyArch achieves speedup by using query-level routing to effectively reduce execution overhead; on top of this, the pipeline further amplifies the gain by overlapping CPU preprocessing with GPU inference. In contrast, MoEfication only introduces logical sparsity, which does not necessarily translate into efficient physical execution on GPUs.

\subsubsection{Peak GPU Memory (VRAM) Consumption}
\label{sec:peak_memory}
\begin{figure}[t]
    \centering
    \begin{subfigure}[t]{0.48\linewidth}
        \centering
        \includegraphics[width=\linewidth]{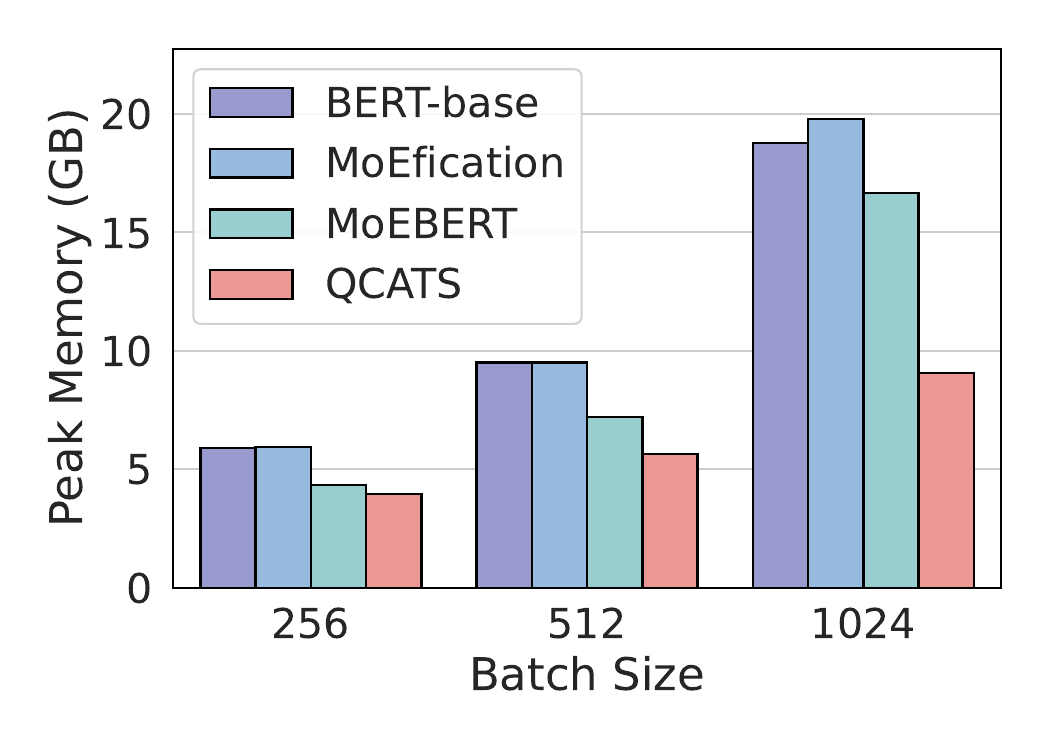}
        \vspace{-2em} 
        \caption{CVE (BERT)}
        \label{fig:mem_bert_cve}
    \end{subfigure}
    \hfill 
    \begin{subfigure}[t]{0.48\linewidth}
        \centering
        \includegraphics[width=\linewidth]{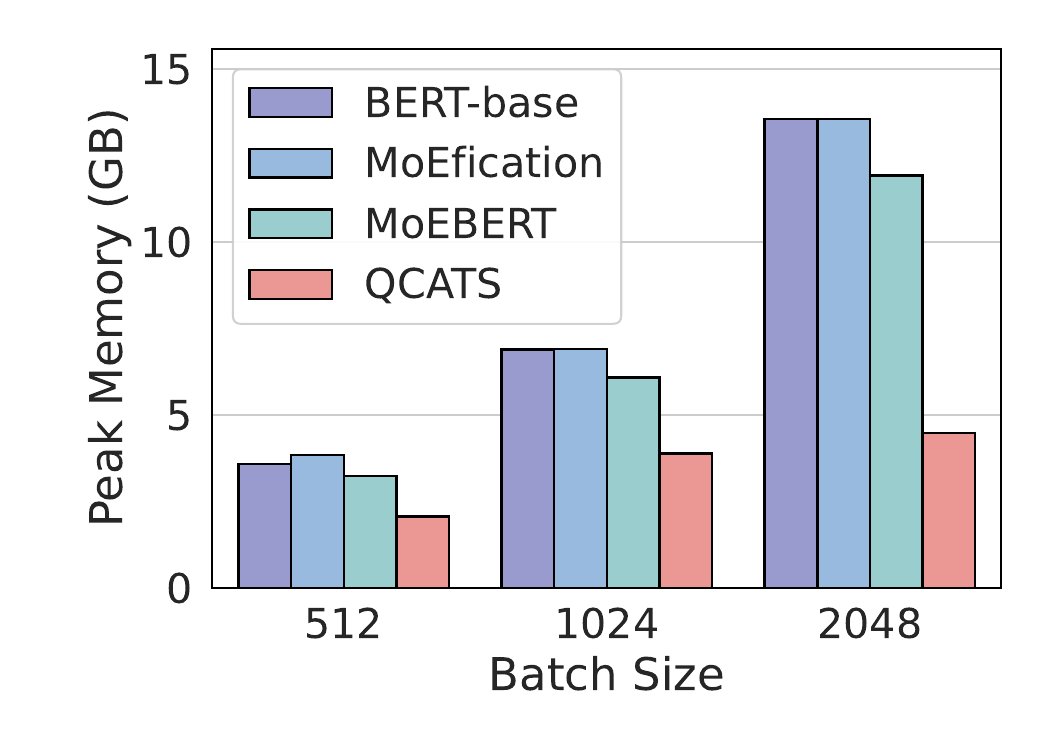}
        \vspace{-2em} 
        \caption{Kickstarter (BERT)}
        \label{fig:mem_bert_ks}
    \end{subfigure} 
    \begin{subfigure}[t]{0.45\linewidth}
        \centering
        \includegraphics[width=\linewidth]{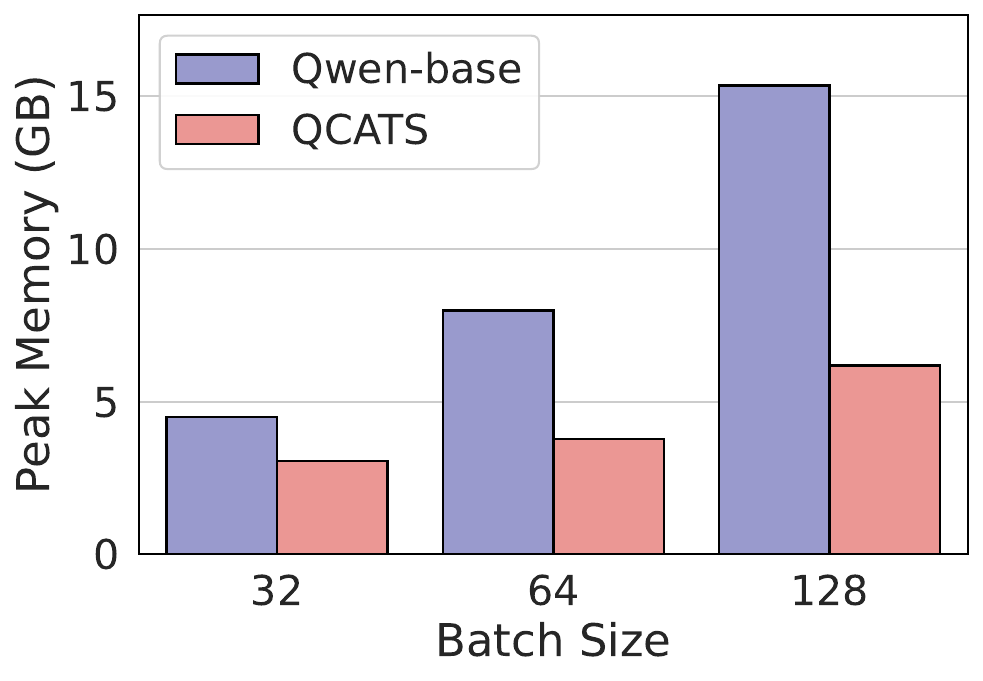}
        \vspace{-2em} 
        \caption{CVE (Qwen)}
        \label{fig:mem_qwen_cve}
    \end{subfigure}
    \hfill 
    \begin{subfigure}[t]{0.45\linewidth}
        \centering
        \includegraphics[width=\linewidth]{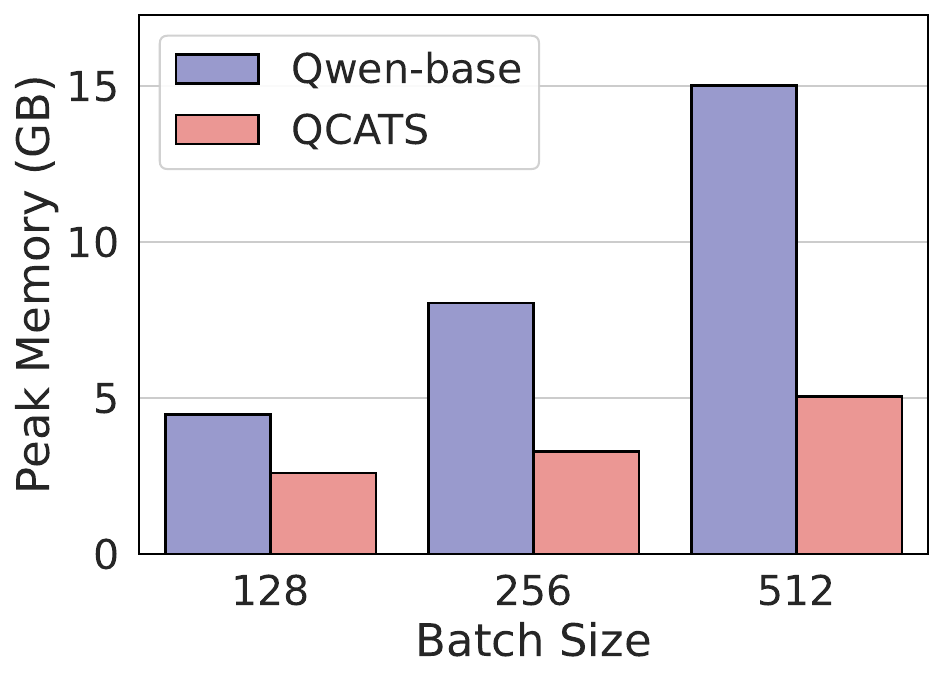}
        \vspace{-2em} 
        \caption{Kickstarter (Qwen)}
        \label{fig:mem_qwen_ks}
    \end{subfigure} 
    \caption{Peak Memory usage comparison.}
    \label{fig:memory_efficiency}
\end{figure}

As shown in Fig.~\ref{fig:memory_efficiency}, we evaluate peak GPU memory usage under the \textit{efficiency evaluation} setting (\S\ref{sec:experimental}). Given that the static weights of BERT and Qwen occupy only about 220 MB and 1.2 GB, respectively, the substantial reduction in peak memory usage achieved by \MyArch indicates that its memory savings mainly come from reduced activation memory. \MyArch allocates memory only to the activated expert subsets and reduces padding through bucketing, thereby effectively compressing the runtime memory footprint.


\subsection{System Effectiveness (RQ2)}
\label{subsec:effectiveness}


\begin{table}[t]
    \centering
    \caption{Comparison of Accuracy and Weighted F1-Score. \textbf{Boldface} indicates the best performance.}
    \label{tab:main_results}
    \resizebox{\columnwidth}{!}{
    \begin{tabular}{l c c c c}
    \toprule
    {\textbf{Model}} & \textbf{Amazon} & \textbf{Kickstarter} & \textbf{OSHA} & \textbf{CVE} \\
     & \small{(Acc / F1)} & \small{(Acc / F1)} & \small{(Acc / F1)} & \small{(Acc / F1)} \\
    \midrule
    BERT-base & 89.7 / 88.6 & 81.4 / 81.0 & \textbf{93.0 / 92.9} & 77.1 / 77.0 \\
    MoEfication & 89.5 / 88.3 & 80.8 / 80.4 & 92.7 / 92.4 & 76.7 / 76.4 \\
    MoEBERT & \textbf{89.9 / 88.7} & 81.6 / 81.3 & \textbf{93.0 / 92.9} & \textbf{78.1 / 78.0} \\
    \textbf{\MyArch(BERT)} & 89.6 / 88.5 & \textbf{81.8 / 81.4} & \textbf{93.0 / 92.9} & 77.9 / 77.8 \\
    \midrule
    Qwen-base & 89.2 / \textbf{87.5} & 80.1 / 80.0 & 92.2 / 91.9 & 74.1 / 73.7 \\
    \textbf{\MyArch(Qwen)} & \textbf{89.5 / 87.5} & \textbf{81.2 / 81.0} & \textbf{92.8 / 92.6} & \textbf{76.6 / 76.2} \\
    \bottomrule
    \end{tabular}
    }
\end{table}
To answer \textbf{RQ2}, we evaluate the prediction robustness of \MyArch under the \textit{effectiveness evaluation} setting (\S\ref{sec:experimental}) using accuracy and weighted F1-score.
Table~\ref{tab:main_results} presents the comparative performance of all models across four datasets. 
It demonstrates that QCATS preserves task accuracy while activating only a subset of FFN parameters.

\MyArch consistently rivals or outperforms dense baselines and matches the performance of MoEBERT, despite using a much simpler routing strategy. 
Our results show that for database-oriented tasks, understanding the global context of the entire query is sufficient to select the right experts. This implies that the heavy computational cost of routing every token is redundant, as \MyArch delivers the same level of accuracy with a much more efficient, query-level approach.

\subsection{Ablation \& Breakdown (RQ3)}
\label{sec:ablation}

This section answers \textbf{RQ3} by evaluating the specific impact of key components.
We perform a latency breakdown to evaluate the impact of the sparsity mechanism of \MyArch and the hierarchical bucketing.
Then, we conduct an ablation study to isolate the roles of the routing mechanism and knowledge distillation in maintaining model accuracy.

\subsubsection{Latency Breakdown and Bottleneck Analysis}
\label{sec:breakdown_analysis}
\begin{figure}[!t]
    \centering
    \captionsetup[subfigure]{skip=2pt} 

    \includegraphics[width=0.95\linewidth]{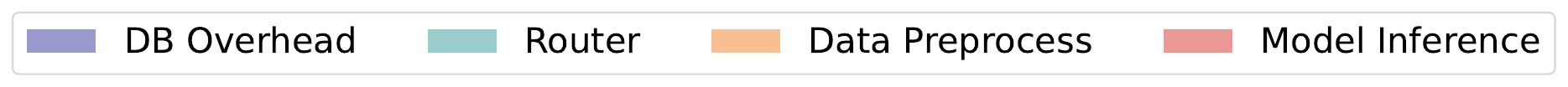}
    \vspace{0.2em} 

    \begin{subfigure}[t]{0.49\linewidth}
        \centering
        \includegraphics[width=\linewidth]{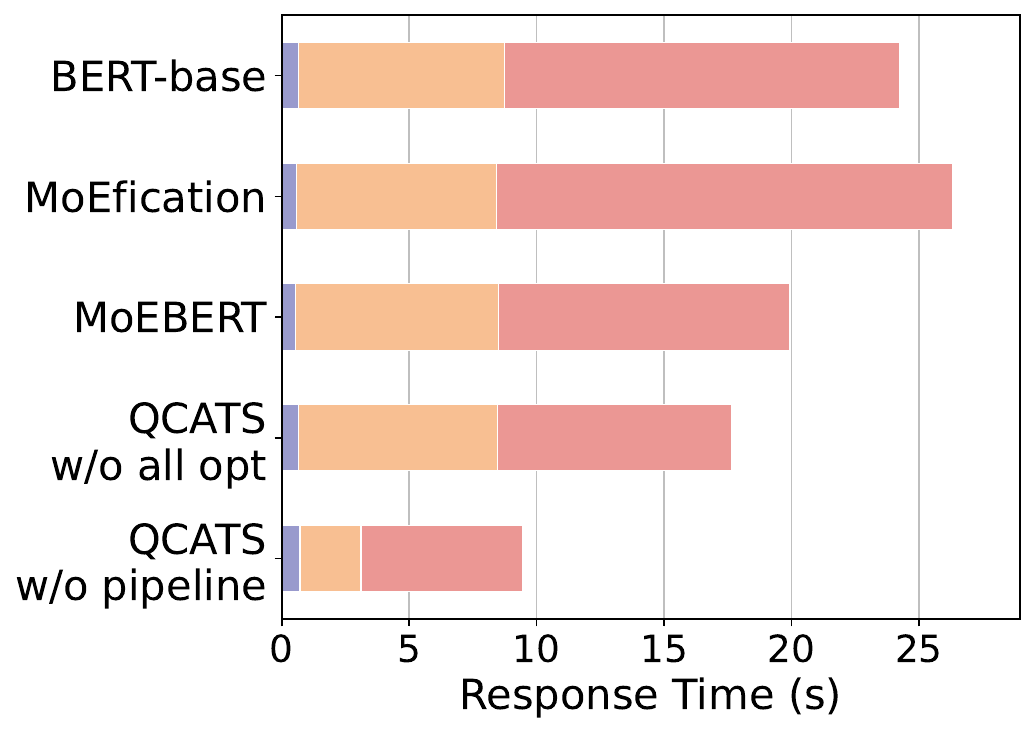}
        \caption{Amazon (BERT)}
        \label{fig:breakdown_a}
    \end{subfigure}
    \hfill
    \begin{subfigure}[t]{0.49\linewidth}
        \centering
        \includegraphics[width=\linewidth]{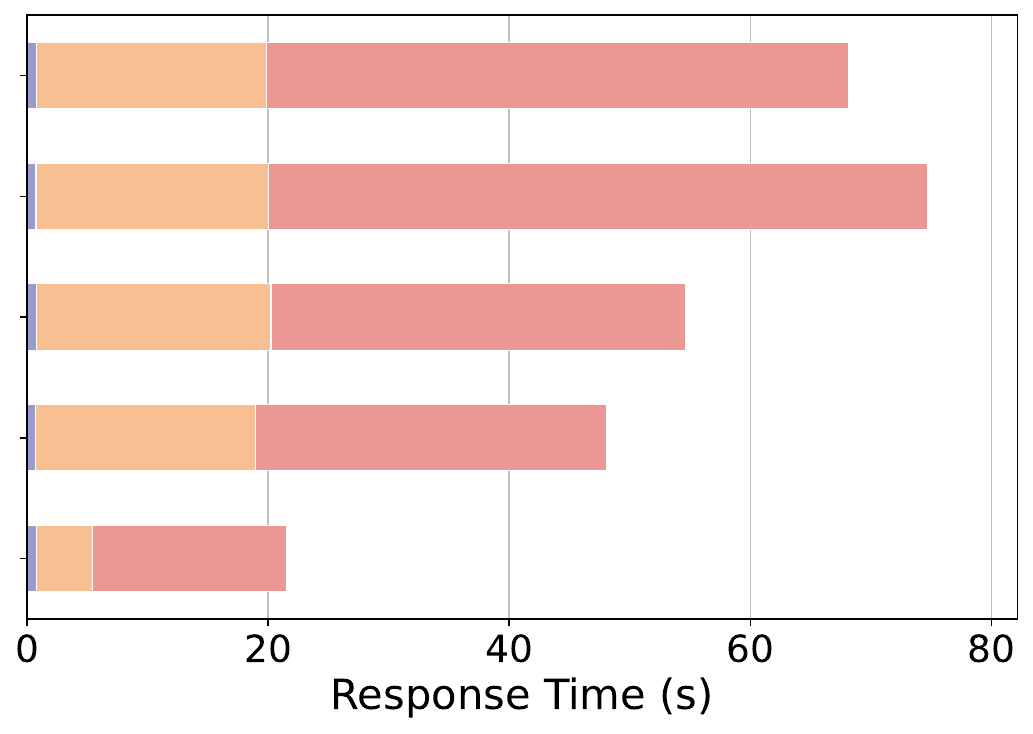}
        \caption{OSHA (BERT)}
        \label{fig:breakdown_b}
    \end{subfigure}

    \vspace{0.2em} 

    \begin{subfigure}[t]{0.49\linewidth}
        \centering
        \includegraphics[width=\linewidth]{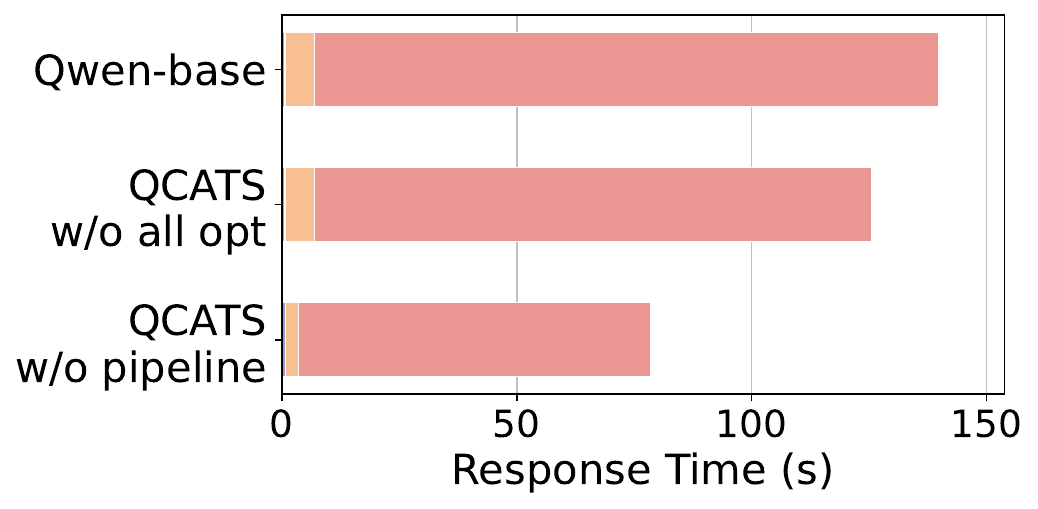}
        \caption{Amazon (Qwen)}
        \label{fig:breakdown_c}
    \end{subfigure}
    \hfill
    \begin{subfigure}[t]{0.49\linewidth}
        \centering
        \includegraphics[width=\linewidth]{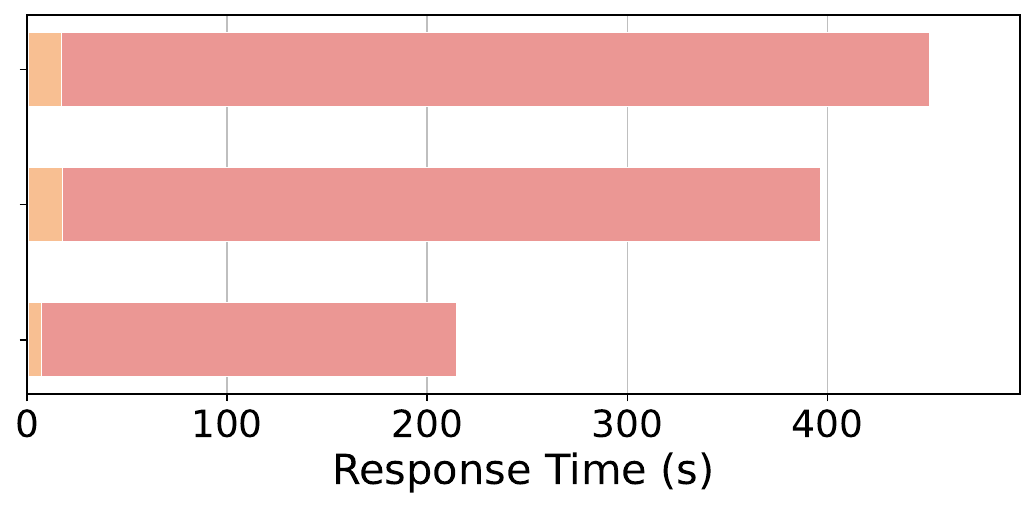}
        \caption{OSHA (Qwen)}
        \label{fig:breakdown_d}
    \end{subfigure}

    \vspace{-0.5em} 
    \caption{Latency Breakdown.}
    \label{fig:breakdown}
\end{figure}
To analyze the sources of \MyArch's performance gains in database scenarios, we executed a query that retrieves 100,000 rows and decomposed the total response time into four main components: database overhead, router computation (specific to \MyArch), data preprocessing, and model inference. As shown in Fig.~\ref{fig:breakdown}, the \MyArch router (green) introduces only negligible overhead. During model inference, \MyArch not only reduces latency by lowering FLOPs through sparsification, but also avoids the runtime overhead of fine-grained dynamic routing in MoE baselines through query-level global routing. Meanwhile, by reducing padding through hierarchical bucketing, \MyArch w/o pipeline outperforms \MyArch w/o all opt in both preprocessing and inference. Finally, the latency breakdown further confirms the necessity of pipelining for lightweight models such as BERT: once GPU inference latency is significantly reduced, CPU preprocessing remains a non-negligible bottleneck.



\subsubsection{Context-Aware Router and Knowledge Distillation}

\begin{figure}[t]
    \centering
    \captionsetup[subfigure]{skip=2pt} 
    \includegraphics[width=1.0\linewidth]{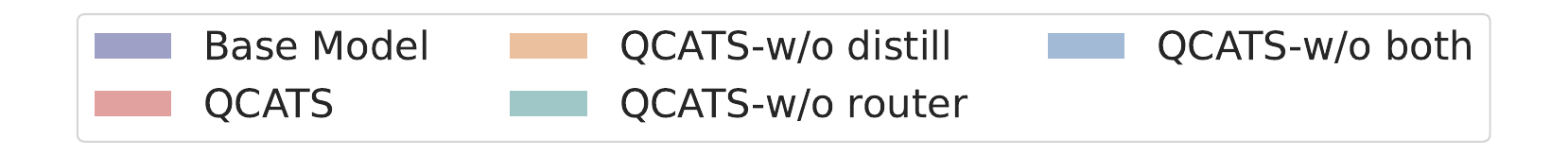}
    \vspace{1.0em} 
    \begin{subfigure}[t]{0.49\linewidth}
        \centering
        \includegraphics[width=\linewidth]{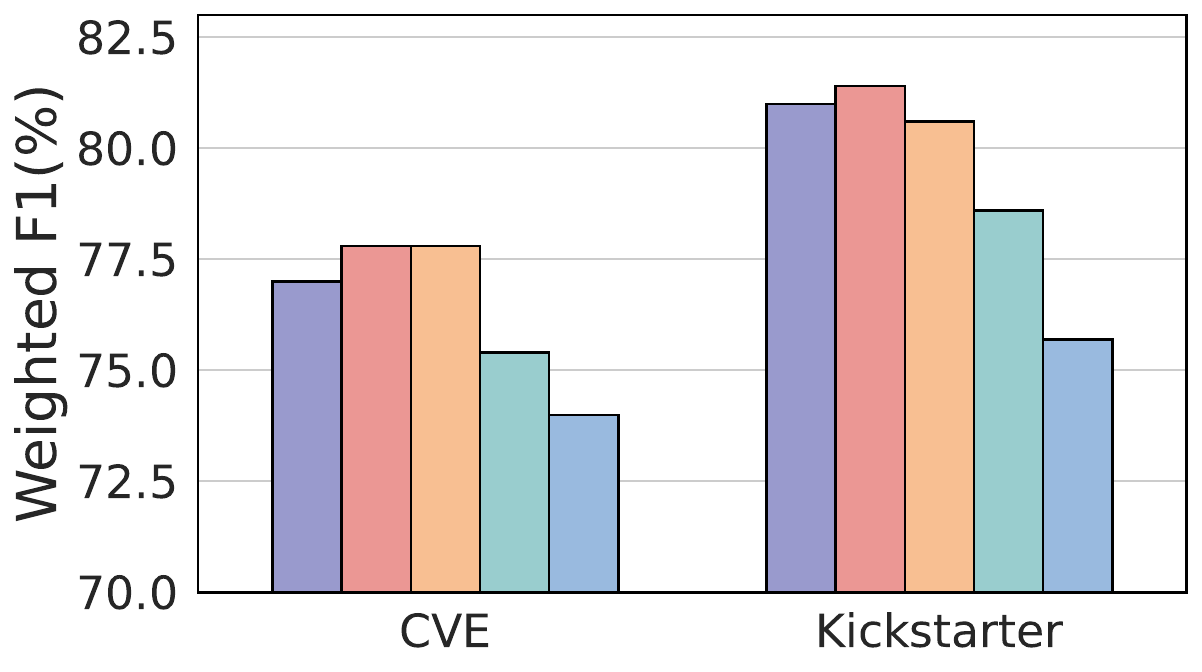}
        \caption{BERT Arch}
        \label{fig:abl_bert}
    \end{subfigure}
    \hfill
    \begin{subfigure}[t]{0.49\linewidth}
        \centering
        \includegraphics[width=\linewidth]{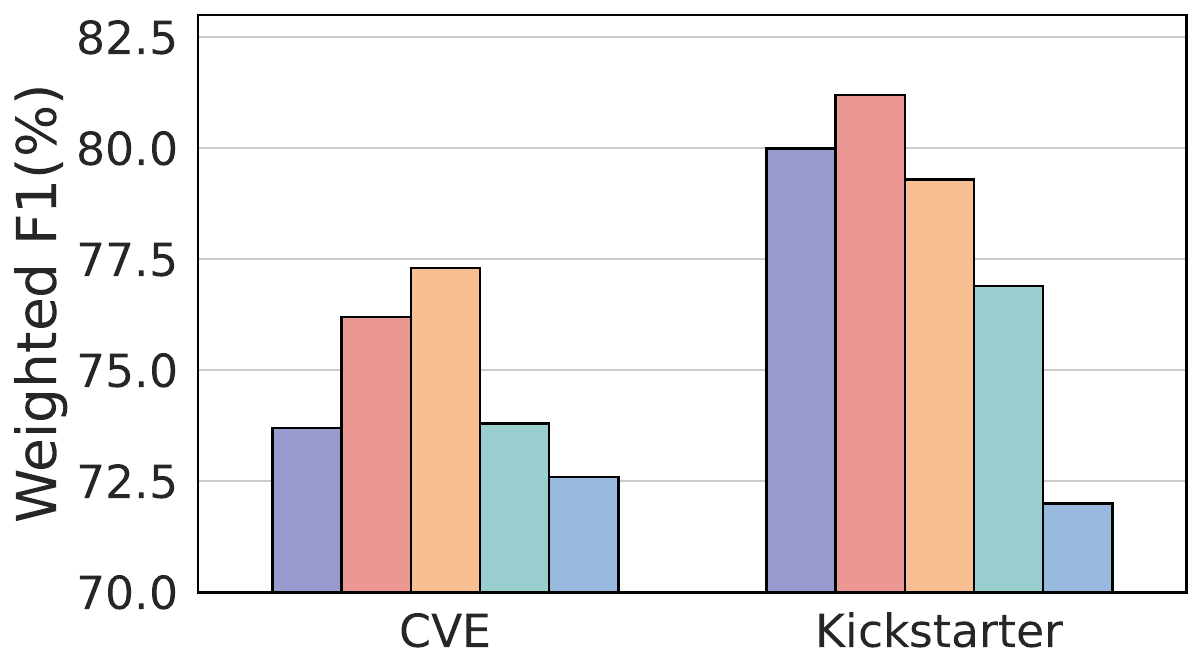}
        \caption{Qwen Arch}
        \label{fig:abl_qwen}
    \end{subfigure}

    \vspace{-1em} 
    \caption{Ablation study.}
    \label{fig:ablation}
\end{figure}
To quantify the independent contributions of knowledge distillation (KD) and dynamic routing, we compare four ablation variants under the \textit{effectiveness evaluation} setting (Fig.~\ref{fig:ablation}). The results show that \MyArch variants with routing (red/orange) consistently achieve performance comparable to the dense baseline, while the \textit{w/o distill} variant even surpasses the teacher model on the Qwen-CVE task, indicating that KD is not the decisive factor behind the performance gain. In contrast, removing routing leads to a clear performance drop, showing that the routing mechanism is critical for expert specialization.

\subsection{Sensitivity \& Robustness (RQ4)}
\label{sec:activation_analysis}

To answer \textbf{RQ4}, we study sensitivity to expert granularity and context columns, evaluate diverse SQL workloads and test transferability to a similar domain.

\subsubsection{Expert granularity}
\begin{figure}[!t]
    \centering
    \begin{subfigure}[!t]{0.45\linewidth}
        \centering
        \vspace{1em}
        \includegraphics[width=\linewidth]{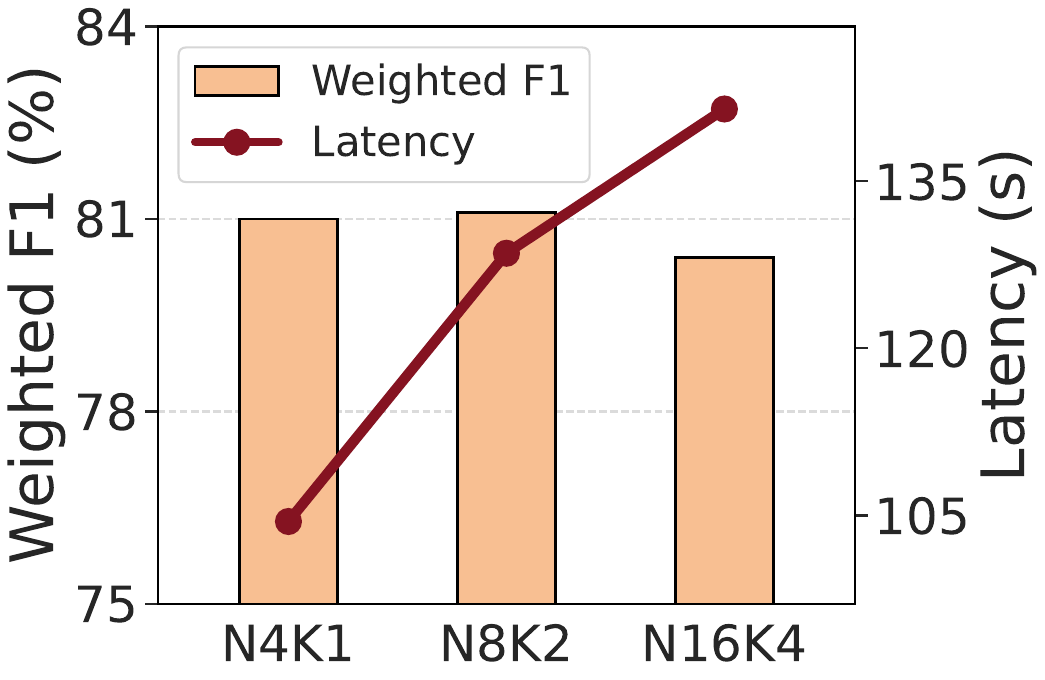}
        \caption{Impact of expert granularity ($N$ and $K$).}
        \label{fig:hyperparam}
    \end{subfigure}
    \hfill
    \begin{subfigure}[!t]{0.53\linewidth}
        \centering
        \includegraphics[width=\linewidth]{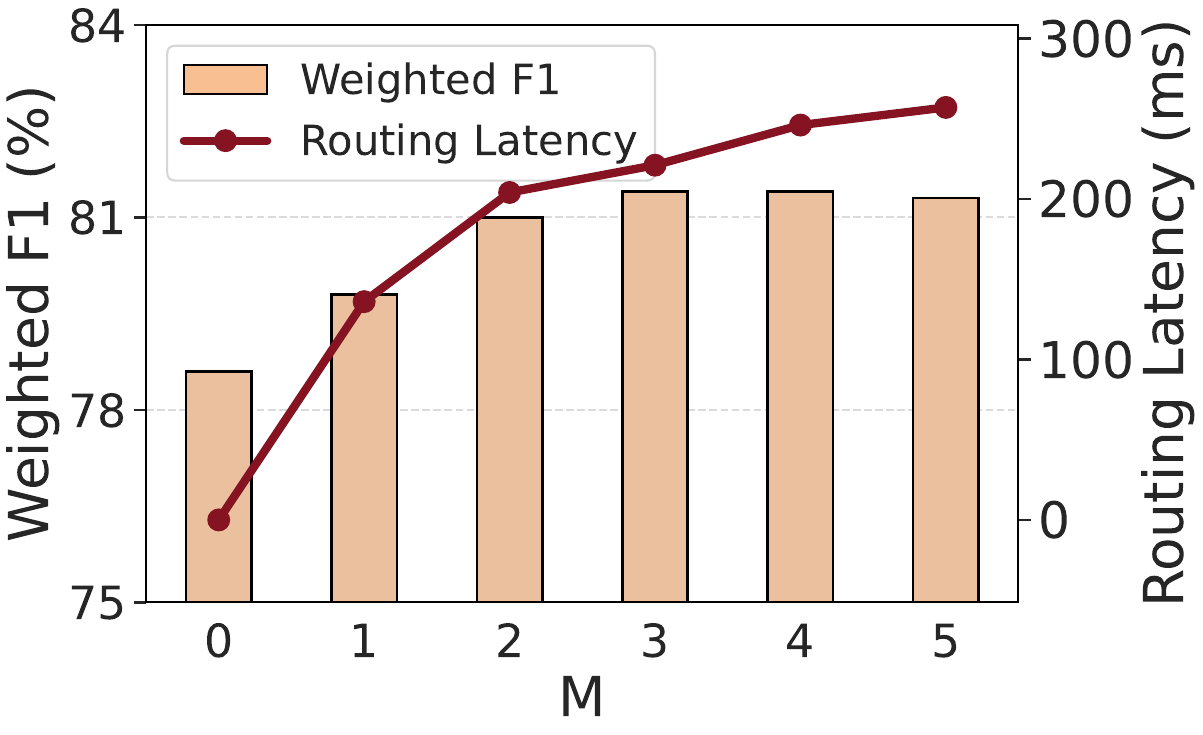}
        \vspace{-2em}
        \caption{Impact of context column.}
        \label{fig:context}
    \end{subfigure}
    \caption{Hyperparameter experiments.}
    
\end{figure}
Following the \textit{effectiveness evaluation} setting, we investigated expert granularity ($N$ vs. $K$) on the Kickstarter dataset while maintaining a constant 25\% activation ratio. Fig.~\ref{fig:hyperparam} reveals high performance stability, indicating that accuracy depends primarily on parameter retention rather than slicing size.
However, the latency for a query retrieving 100,000 rows, highlights the cost of finer granularity. 
Therefore, we adopt N4K1 as the default to maximize computational density and throughput without compromising predictive performance.

\subsubsection{Context Columns}
Under the \textit{effectiveness evaluation} setting, we analyze the sensitivity of the retained context column count $M$ on the Kickstarter dataset. Fig.~\ref{fig:context} illustrates this trade-off between weighted F1-score and average routing latency across varying $M$. As shown, a  small $M$ fails to capture the global data distribution, weakening context alignment and routing accuracy. Conversely, a large $M$ introduces redundant attributes; this increases preprocessing overhead.

\begin{figure*}[!t]
    \centering
    \includegraphics[width=\linewidth]{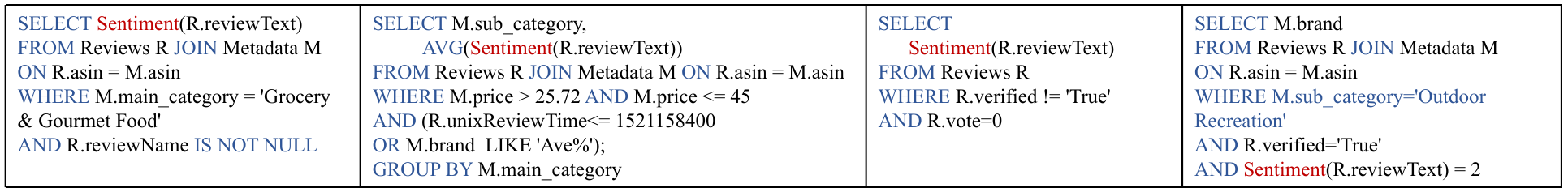}
    
    \vspace{0.5em} 
    
    \begin{minipage}{0.24\linewidth}
        \centering (a) Q1: Normal
    \end{minipage}
    \hfill
    \begin{minipage}{0.24\linewidth}
        \centering (b) Q2: Aggregation
    \end{minipage}
    \hfill
    \begin{minipage}{0.24\linewidth}
        \centering (c) Q3: No Context
    \end{minipage}
    \hfill
    \begin{minipage}{0.24\linewidth}
        \centering (d) Q4: UDF Predicate
    \end{minipage}
    
    \vspace{0.2em}
    
    \caption{Predictive queries in the experiment.}
    \label{fig:query}
\end{figure*}

\begin{figure}[!t]
    \centering
    \begin{subfigure}{0.49\linewidth}
        \centering
        \includegraphics[width=\linewidth]{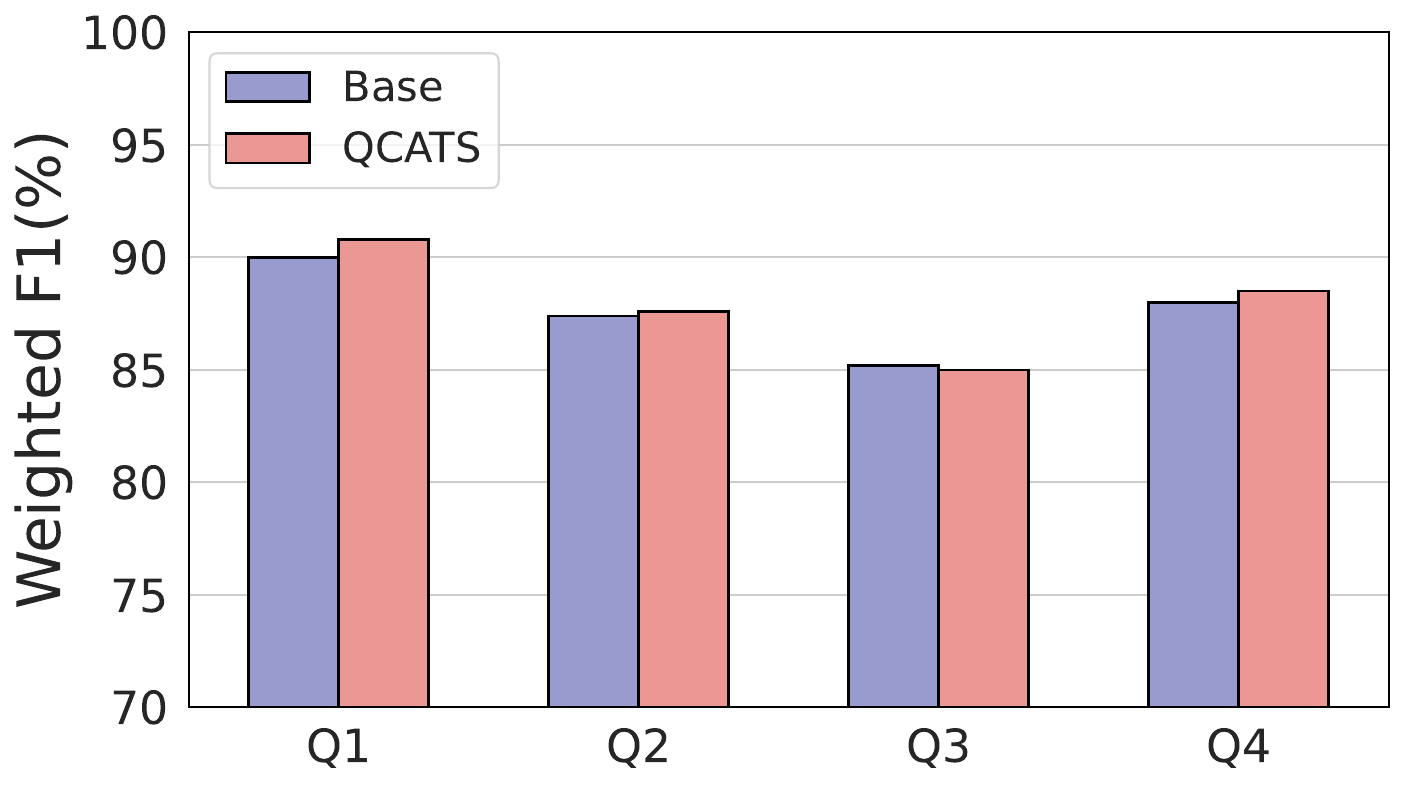}
        \caption{Weighted F1-score.}
        \label{fig:f1_comparison}
    \end{subfigure}
    \hfill 
    \begin{subfigure}{0.49\linewidth}
        \centering
        \includegraphics[width=\linewidth]{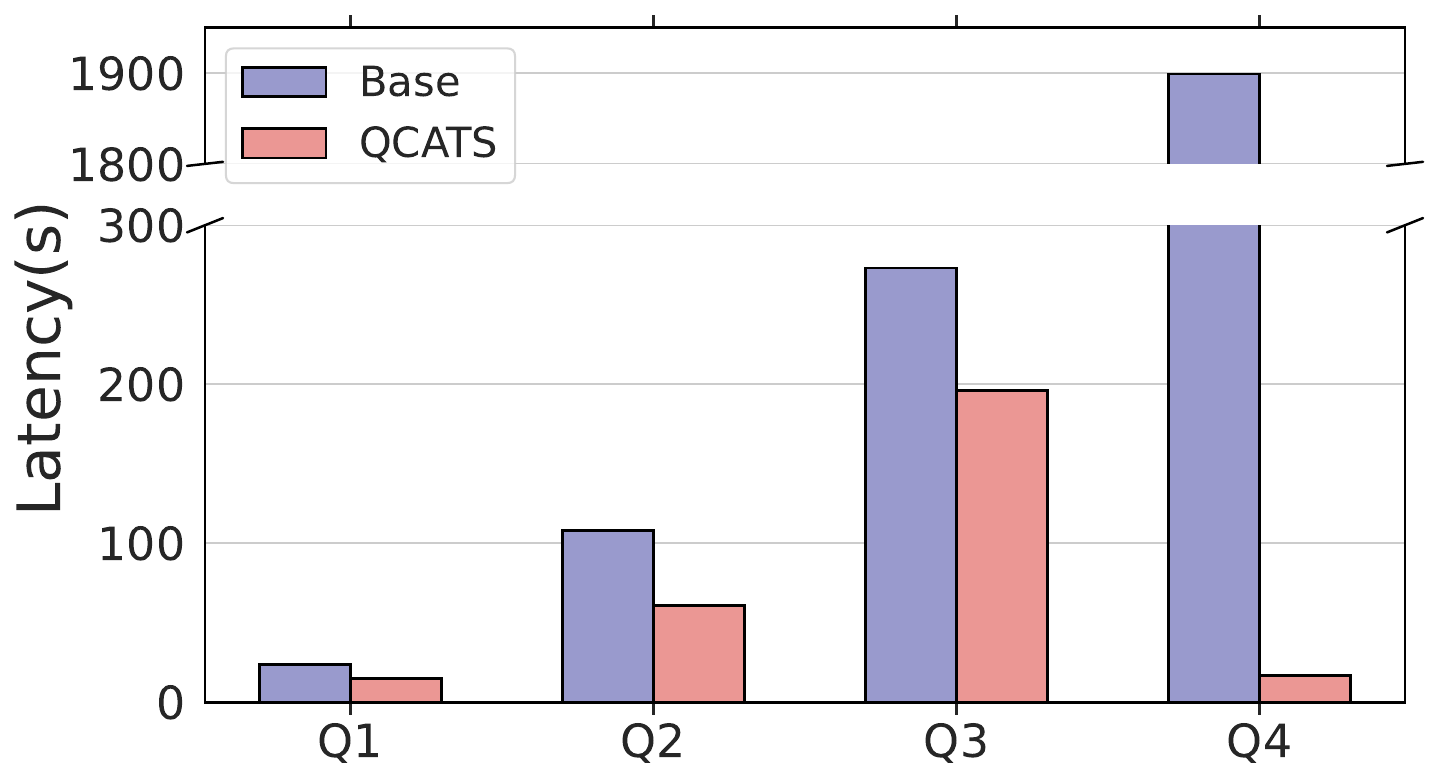}
        \caption{End-to-end latency.} 
        \label{fig:latency_comparison}
    \end{subfigure}

    \caption{Performance evaluation of QCATS across diverse SQL workloads.}
    \label{fig:overall_performance}
\end{figure}


\subsubsection{QCATS under diverse SQL workloads}
To evaluate the adaptability of QCATS across diverse SQL workloads, we analyze four representative queries on the Amazon dataset using the Qwen backbone (Fig.~\ref{fig:query}). As shown in Fig.~\ref{fig:overall_performance}, QCATS consistently reduces end-to-end latency across all scenarios while exerting only minimal impact on prediction accuracy. Notably, even for Q4, where the UDF appears in the filter predicate, QCATS still achieves substantial latency reduction by significantly decreasing the amount of data fed into the model through query rewriting.

\subsubsection{Case Study: Transferability to Similar Domains}
\label{subsec:transferability}
\begin{table}[t]
    \centering
    \caption{Transferability to Similar Domains. \textbf{Boldface} indicates the best performance.}
    \label{tab:transferability}
    \resizebox{\columnwidth}{!}{
    \begin{tabular}{l c c }
    \toprule
    \multirow{2}{*}{\textbf{Model}} & \textbf{Source Domain (Amazon)} & \textbf{Target Domain (Flipkart)}  \\
     & \small{(Acc / F1)} & \small{(Acc / F1)}  \\
    \midrule
    BERT-base & \textbf{89.7 / 88.6} & \textbf{88.7} / 86.5 \\
    \textbf{\MyArch(BERT)} & 89.6 / 88.5 & \textbf{88.7 / 86.6} \\
    \midrule
    Qwen-base & 89.2 / \textbf{87.5} &  88.6 / \textbf{85.6}\\
    \textbf{\MyArch(Qwen)} & \textbf{89.5 / 87.5} & \textbf{88.8 / 85.6} \\
    \bottomrule
    \end{tabular}
    }
\end{table}
To assess generalization, we directly transfer models trained on Amazon to the Flipkart dataset~\cite{tharunmss_water_bottle, guhan_speaker_sentiments, maddali_laptop_reviews} in a zero-shot manner, using \texttt {Product Category} and \texttt {Product Price} as routing contexts (following \textit{effectiveness evaluation} setting). As shown in Table~\ref {tab:transferability}, \MyArch achieves performance comparable to dense baselines in this similar-domain setting, indicating that its query-based slicing mechanism preserves both transferability and core model capabilities.

\section{Related Work}
\label{sec:related}

\noindent
\lyy{\textbf{In-database ML and predictive query processing}} have long focused on integrating model inference into SQL execution pipelines. 
Early approaches~\cite{hellerstein2012madlib, boehm2020systemds} and modern Cloud APIs~\cite{databricks2024serving, google2024inference} predominantly rely on UDFs to invoke models, but they treat inference as a black-box task, discarding rich query-time context. 
To enable deeper optimization, some frameworks~\cite{park2022end,lin2023smartlite} decompose models into basic relational or tensor operators to perform fine-grained execution. 
However, these white-box optimizations are largely confined to traditional ML or neural networks and do not scale to the massive parameter space of Transformers. 
In contrast, our work bridges this gap: we move beyond external data movement or scheduling by coupling database analytics with model internals. 
By leveraging query context, we enable the deterministic customization of the model inference path itself for \yy{Transformer-based models}.

\noindent
\lyy{\textbf{Database-aware model serving}} has been explored to improve inference flexibility, but existing approaches differ fundamentally from \MyArch in optimization granularity and system objectives.
ARM-Net~\cite{cai2021arm} adapts feature interactions at the instance level to improve prediction accuracy, without targeting inference efficiency or execution-path optimization.
Model Slicing~\cite{cai2019model} supports elastic inference cost via width-wise slicing driven by resource budgets, but does not exploit query semantics or workload locality.
\yy{
LEADS~\cite{zeng2024powering} also exploits database context, but selects among replicated experts or multiple conventional models primarily to improve prediction quality for structured-data analytics. Like the preceding approaches, it does not specifically address Transformer-based text inference. QCATS instead uses query context to deterministically select FFN neuron slices within a single Transformer backbone. This targets text prediction while reducing inference cost and GPU memory usage and preserving dense-model accuracy.
}
\yy{Furthermore, all the aforementioned methods do not specifically address the inference characteristics of Transformer-based text prediction, while \MyArch uses query context to deterministically select model parameters, thereby improving inference efficiency while preserving accuracy.
}

\noindent
\textbf{\lyy{Efficient Transformer inference and MoE}} architecture~\cite{lepikhin2020gshard, fedus2022switch} scales model capacity while maintaining manageable inference costs. Recent efforts in \textit{MoE-fication}~\cite{zhang2022moefication, zhang2024exploring, zuo2022moebert, cai2024textit} have extended this concept by refactoring pre-trained dense models into expert-based structures. However, standard MoEs rely on token-level, layer-wise routing, which treats input data as independent instances and ignores high-level semantic contexts. While recent works attempt to optimize routing via decoupling~\cite{hwang2024pre,cai2024textit} or speculation~\cite{du2024sida, he2024expertflow}, they remain constrained by micro-level token dependencies, limiting optimization in data-intensive database environments.
In contrast, \MyArch is tailored for in-database analytics, where data is accessed via structured queries rather than random batches. We exploit the query context to implement query-level global routing. By leveraging these database-specific hints to deterministically pre-select expert slices, we eliminate runtime dispatching overhead and enable efficient, homogeneous batching that is unattainable in context-agnostic systems.

\section{Conclusion}
\label{sec:conclusion}
We presented \MyArch, a query context-aware transformer slicing framework that enables efficient sparse inference for in-database predictive queries.
By shifting conditional execution from token level to query level, \MyArch aligns sparse neural execution with database semantics.
Extensive experimental results demonstrate that \MyArch significantly improves inference throughput and reduces memory consumption while preserving predictive accuracy compared to dense baselines.
This suggests that query-aware execution provides a practical foundation for bridging \yy{Transformer-based} model inference with high-performance database systems.

We believe this work opens new opportunities for designing database-native inference engines that treat neural execution as a first-class query operator rather than an external black-box service.
\yy{
Building on this direction, we will evaluate QCATS on larger Transformer backbones and more complex, realistic workloads under distribution shifts and heterogeneous domains.
We will also investigate adaptation and cross-domain transferability of expert libraries and routing policies, while interactions with generative-model serving systems remain separate future work.
}

\section{Acknowledgment}
\yy{
This work was supported by the Fundamental and Interdisciplinary Disciplines Breakthrough Plan of the Ministry of Education
of China (JYB2025XDXM103), the National Natural Science Foundation of China (Grant No. 62602571), and the Leading Talent of Technological Innovation Program (No. 2023R5214) of Zhejiang Province.
}

We disclose that GPT was used only to assist with English polishing, grammar improvement, and the writing of some utility code. The research questions, overall design, implementation process, experiments, analysis of results, and conclusions were developed entirely by the authors.

\putbib[references]

\end{bibunit}

\appendices
\begin{bibunit}
\section*{APPENDIX}


We attach the analysis and implementation details of QCATS in this appendix.

\section*{Query Context Analysis}

To evaluate the neuron activation across diverse query contexts, we constructed a benchmark of 16 predictive queries based on the Kickstarter dataset. We first selected 8 seed queries ($Q1$-$Q8$) from the workload and applied minor perturbations—such as adding, deleting, or modifying filtering conditions—to generate a corresponding set of variants ($Q9$-$Q16$).

\begin{table}[h] 
    \centering
    \caption{Filter Predicates for the Queries.}
    \label{tab:query_predicates}
    \small 
    \renewcommand{\arraystretch}{1.2} 
    
    \begin{tabularx}{\linewidth}{c X} 
        \toprule
        \textbf{ID} & \textbf{Filter Predicates $\mathcal{P}$} \\
        \midrule
        
        Q1 & \texttt{main\_category = `technology' AND sub\_category = `Gadgets'} \\
        Q2 & \texttt{sub\_category = `Video Games'} \\
        Q3 & \texttt{main\_category = `music' AND 1000 < goal < 2000} \\
        Q4 & \texttt{main\_category = `publishing' AND country = `GB' AND 20 < duration < 30} \\
        Q5 & \texttt{main\_category = `film \& video' AND sub\_category = `Drama'} \\
        Q6 & \texttt{sub\_category = `Drinks'} \\
        Q7 & \texttt{main\_category = `fashion' AND goal < 2000} \\
        Q8 & \texttt{main\_category = `art'} \\
        
        \midrule 
        
        Q9 & \texttt{main\_category = `technology' AND sub\_category = `Gadgets' AND city = `Hong Kong'} \\
        Q10 & \texttt{sub\_category = `Video Games' AND country = `FR'} \\
        Q11 & \texttt{main\_category = `music' AND goal < 2000 AND duration > 30} \\
        Q12 & \texttt{main\_category = `publishing' AND country = `CA'} \\
        Q13 & \texttt{sub\_category = `Drama' AND city = `London'} \\
        Q14 & \texttt{sub\_category = `Vegan'} \\
        Q15 & \texttt{main\_category = `fashion' AND goal > 2000} \\
        Q16 & \texttt{sub\_category = `Sculpture' AND currency = `GBP'} \\
        
        \bottomrule
    \end{tabularx}
\end{table}

All predictive queries follow a unified structure, formally defined as:

\begin{equation}
\begin{split}
&\texttt{\textbf{SELECT} MODEL(name\_blurb)} \\
&\texttt{\textbf{FROM} Kickstarter} \\
&\texttt{\textbf{WHERE} } \mathcal{P};
\end{split}
\end{equation}

\noindent where $\mathcal{P}$ represents the set of variable predicates detailed in Table~\ref{tab:query_predicates}.

Given that $Q9$-$Q16$ are derivatives of $Q1$-$Q8$, their result sets often exhibit inclusion relationships (e.g., a modified query's results being a subset of the original). To decouple the analysis of neuron activation patterns from inherent data overlap, we randomly sampled 200 distinct tuples from the result set of each query rather than using the full retrieval sets. We then monitored the activation states of a fine-tuned dense model on these samples. Aligned with our experimental sparsity setting—which maintains a 25\% activation rate—we focused our analysis exclusively on the top 768 neurons (the top 25\% of the post-activation hidden layer) ranked by average magnitude.

\begin{figure}[h] 
    \centering
    \includegraphics[width=1.0\linewidth]{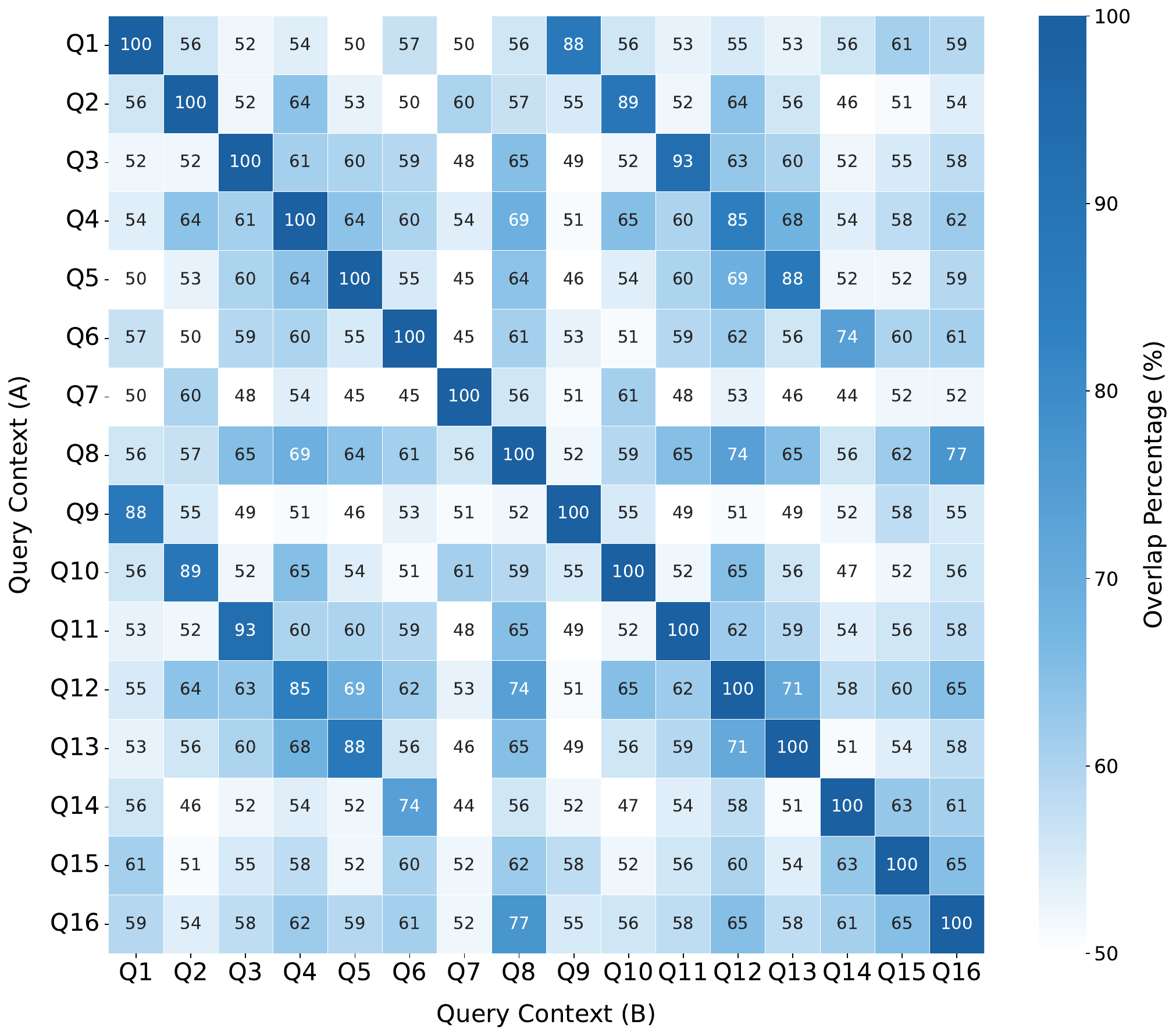}
    
    \caption{\textbf{Cross-Query Top-768 Neuron Overlap Matrix.} 
    This heatmap visualizes the pairwise overlap percentage of the top-768 most active neurons across 16 distinct query contexts ($Q1$-$Q16$). 
    Darker blue indicates higher overlap (similarity), while lighter colors indicate distinct activation patterns. 
    Diagonal elements are 100\% by definition.}
    
    \label{fig:overlap}
    
    \vspace{-1em} 
\end{figure}

The pairwise overlap rates of these top-ranked neurons are visualized in Fig.~\ref{fig:overlap}. We observe strong context locality: there is significant divergence between unrelated groups (e.g., between the $Q1$-$Q8$ block and unrelated queries in $Q9$-$Q16$), whereas semantically related pairs (e.g., $Q1$ \& $Q9$, $Q2$ \& $Q10$) exhibit high similarity. This empirical evidence supports our core hypothesis: different query contexts induce distinct activation patterns in Transformer feed-forward networks (FFNs).

\section*{hyperparameter settings}
In \textit{Stage I: Expert Construction}, we first employ the KNN algorithm to partition the dataset into $N=4$ semantic groups. Subsequently, we sample $\mathcal{D}_n=500$ representative instances from each group to perform activation profiling. This sample size is chosen to mitigate statistical noise while ensuring robust neuron distinction, thereby maintaining high profiling accuracy with minimal offline computational overhead. 

Based on this grouping, we adopt an $N4K1$ configuration, where the feed-forward network is partitioned into $N=4$ experts, and $K=1$ expert is dynamically activated during inference. This results in a fixed activation rate of 25\%, consistent with settings established in prior work~\cite{zuo2022moebert,zhang2022moefication}. We prioritize the $N=4, K=1$ configuration because finer-grained splitting (e.g., $N=8, K=2$ or $N=16, K=4$), while potentially allowing better collaboration among semantic experts, incurs significant inference latency and introduces load balancing challenges while yielding only limited accuracy gains. 

Furthermore, we designate 50\% of the neurons as shared neurons $U_{\text{shared}}$ and the remaining 50\% as context-specific neurons. This split is motivated by the observation in Fig.~\ref{fig:overlap}, which indicates that the overlap of highly activated neurons across varying query contexts is approximately 50\%.

For \textit{One-time Feature Selection} in \textit{Stage II}, we train a Random Forest classifier on a subset of 30,000 samples to evaluate metadata importance, selecting the top $M=3$ columns as effective context attributes. 
Empirically, we find that selecting the top three features is sufficient for the datasets used in this work, although the optimal $M$ may vary across different datasets.

In \textit{Context-Aware Expert Routing}, the router is implemented as a Transformer block with a hidden dimension of 512 and 4 attention heads, followed by a 3-layer MLP with a dimension of 256.

During \textit{Phase I: Router Semantic Alignment} of \textit{Decoupled Training Strategy}, we generate ground-truth labels using a calibration coefficient $\eta=0.02$ and a softmax temperature $\tau=1.0$. We selected $\eta=0.02$ based on observations from the training set: this value ensures that the distribution of selected expert indices aligns well with the ranking of expert scores, while simultaneously preventing load imbalance caused by the under-utilization of lower-scoring experts.

Finally, during \textit{Phase II: Context-Driven Joint Expert Training} of \textit{Decoupled Training Strategy}, we set the distillation loss weight to $\alpha=5.0$, consistent with~\cite{zuo2022moebert}.

\putbib[appendix-ref] 
\end{bibunit}
\end{document}